\documentclass{article}
\usepackage{amsfonts,amsthm,amsmath,amssymb}
\usepackage{pdfsync,url}
\usepackage[utf8]{inputenc}
\usepackage{amsmath}

\usepackage[margin=1in]{geometry}
\usepackage{endfloat}

\usepackage{authblk}

\theoremstyle{plain}
\newtheorem{thm}{\protect\theoremname}
\theoremstyle{plain}
\newtheorem{prop}[thm]{\protect\propositionname}
\theoremstyle{definition}
\newtheorem{defn}[thm]{\protect\definitionname}
\theoremstyle{plain}
\newtheorem{cor}[thm]{\protect\corollaryname}

\providecommand{\corollaryname}{Corollary}
\providecommand{\definitionname}{Definition}
\providecommand{\propositionname}{Proposition}
\providecommand{\theoremname}{Theorem}

\usepackage[round]{natbib}

\usepackage{todonotes}
\title{\bf Informed Pooled Testing with Quantitative Assays}

\author[1]{Tao Liu}
\author[1]{Joseph W.\ Hogan}
\author[1]{Wanning Su}
\author[2]{Yizhen Xu}
\author[3]{Michael J.\ Daniels}
\author[4]{Kantor Rami}

\affil[1]{Department of Biostatistics, Center for Statistical Sciences, Brown University School of Public Health, Providence, RI 02912, USA. }
\affil[2]{Department of Biostatistics, Johns Hopkins Bloomberg School of Public Health, Baltimore MD 21218, USA. }
\affil[3]{Department of Statistics, University of Florida, Gainesville, FL 32611, USA.}
\affil[4]{Department of Medicine, the  Alpert Medical School of Brown University, Providence, RI 02912, USA. }

\begin{document}

\maketitle

\vspace{0.8in}

\noindent {\bf Correspondence:}  \\
Tao Liu, PhD (Email: tliu@stat.brown.edu) \\
Associate Professor of Biostatistics \\
Department of Biostatistics\\
Brown University, Box GS-121\\
Providence, RI 02912, USA.  \\

~

\noindent This study was partially supported by NIH/NIAID grants (R01-AI108441, R01-AI136664, P30-AI42853). 

\newpage

\vspace{0.3in}

\begin{abstract}

Pooled testing is widely used for screening for viral or bacterial infections with low prevalence when individual testing is not cost-efficient. Pooled testing with qualitative assays that give binary results has been well-studied. However, characteristics of pooling with quantitative assays were mostly demonstrated using simulations or empirical studies. We investigate properties of three pooling strategies with quantitative assays: traditional two-stage mini-pooling (MP) (Dorfman, 1943), mini-pooling with deconvolution algorithm (MPA) (May et al., 2010), and marker-assisted MPA (mMPA) (Liu et al., 2017). MPA and mMPA test individuals in a sequence after a positive pool and implement a deconvolution algorithm to determine when testing can cease to ascertain all individual statuses. mMPA uses information from other available markers to determine an optimal order for individual testings. We derive and compare the general statistical properties of the three pooling methods. We show that with a proper pool size, MP, MPA, and mMPA can be more cost-efficient than individual testing, and mMPA is superior to MPA and MP. For diagnostic accuracy, mMPA and MPA have higher specificity and positive predictive value but lower sensitivity and negative predictive value than MP and individual testing. Included in this paper are applications to various simulations and an application for HIV treatment monitoring. 

%[100 words version]
%Pooled (group) testing can increase test cost-effectiveness and is valuable for screening diseases with low prevalence. Pooling with quantitative assays is not well studied. We investigate three pooling strategies with quantitative assays: traditional two-stage mini-pooling (MP), mini-pooling with deconvolution algorithm (MPA), and marker-assisted mini-pooling with deconvolution algorithm (mMPA). We derive and compare their statistical properties and diagnostic accuracy. We provide theoretical support for including a deconvolution algorithm and establish the superiority of mMPA to MPA. Implementations of these methods are illustrated by simulations and application. 

\end{abstract}


%\newpage
%%%%%%%%%%%%%%%
%\listoftodos

%{\bf Need immediate attention}

%{\bf Can work on later}

%%%%%%%%%%%%%%%
\newpage

\section{Introduction} \label{sec:intro}

Pooled testing (also known as group testing) can be traced back to 1940's \citep{dorfman1943detection}. It is often used to identify low-prevalence conditions such as rare diseases, infections, and defects. The most commonly-used pooling strategies are a two-stage testing procedure referred to as mini-pooling (MP), wherein samples of equal amount from $K$ individuals are first combined to form a pool, and a single test is conducted on the pool; and at the second stage, individual samples are tested only if the pool test is positive (otherwise all samples are classified as negative). MP and its variants such as pyramid pooling \citep{Bilder2010, Quinn2000} and matrix pooling \citep{van2011pooling, Litvak1994} are valuable particularly when the conditions to be detected are rare and resources such as testing facilities and assays are constrained. There is abundant literature on pooled testing with \emph{qualitative} assays that yield binary, or so-called Boolean, results (e.g.\ yes/no, true/false, and positive/negative) \citep[c.f.][]{Du1999a}. 

Pooling strategies also can be applied to \emph{quantitative} assays that yield a non-negative numerical value, say $V\in [0,\infty)$, where the condition of interest is defined as a result exceeding a threshold $C$, i.e. $V>C$ \citep{May2010, tilghman2011pooled, Tilghman2015, Liu2017JAIDS}. For example, quantitative RT-PCR assay is routinely used for quantification of viral RNA (e.g.\ HIV, Ebola, and COVID) by measuring the number of copies of specific cDNA targets. Pooling with quantitative assays is similar to pooling with qualitative assays, but some fundamental differences exist. One difference is that pooling with qualitative assays can be described using a Boolean summation process. In the ideal case with no measurement errors, one or more positive samples in a pool imply that the pool is positive and vice versa. In contrast to this, pooling with quantitative assays can be described by a convolution process. Because of the dilution effect, a pool is defined as positive if the test on the pool is greater than $C/K$ (instead of $C$). We proceed to the second stage testing if the pool \emph{possibly} contains one failure (e.g.\ one sample has $V>C$ and all the others have $V=0$). This difference suggests that for pooling with quantitative assays, we cannot draw a conclusion that a pool contains at least one failure if the pool tests positive. Moreover, the quantitative nature of assays makes it possible to incorporate a deconvolution algorithm in the second stage of individual testing after a pool tests positive \citep{May2010} and to further incorporate external information that is associated with $V$~\citep{Liu2017JAIDS}. We review these methods in the context of HIV treatment monitoring as follows. 

For people living with HIV and receiving antiretroviral treatment (ART), treatment effectiveness is monitored by HIV viral load (VL) \citep{Hammer:2006}, which measures the number of copies of virus being replicated and circulated inside the human body. An elevated VL (e.g.\ exceeding $C=1,000$ copies/mL) implies that ART is failing to suppress the virus replications, and the patient needs to be switched to the next-line treatment. 
In resource-poor countries where HIV infection is prevalent, ability to carry out VL testing is constrained by lab infrastructure and cost \citep{Rowley2014, MEDICINSSANSFRONTIERES2012}. Routine and global use of VL testing is not feasible, even though the World Health Organization (WHO) has recommended such practices \citep{WHO2014, Piot2015}. Thus, HIV treatment monitoring strategies that make the effective use of limited VL assays are needed \citep{Liu2013, Koller2015, Cerutti2016}. 

\cite{May2010} proposed a pooling method called ``MP with algorithm'' (MPA) for HIV treatment monitoring, which involves two-stage testing as MP but at the second stage (when a pool tests positive), individual samples are tested sequentially with a stopping algorithm. The stopping algorithm is based on the pool result `subtracting' the results of individual samples that have been tested. The remainder recovers partially the virological status of the remaining individual samples; when the remainder is sufficiently low, all remaining samples are concluded as negative. For example, consider a pool containing three samples. If the pool test result is 500, and testing the first sample yields a result of 1300, then it can be readily deduced that the test results of the remaining two samples cannot exceed $500 \times 3 - 1300 = 200$, even if their actual values cannot be determined. If a cutoff $C>200$ is used, the remaining two samples can be classified as negative without testing. Properties of MPA has been demonstrated using Monte Carlo and clinical studies \citep{tilghman2011pooled, Tilghman2015, Schooley:2007, Kim2014}, 

Building on May et al's method, we proposed another pooling strategy called ``marker-assisted MPA'' (mMPA), which incorporates information from one or more markers that are correlated with $V$ into testing procedure \citep{Liu2017JAIDS}. The marker values are used to determine the risk of failure and following a positive pool,  individual samples are tested sequentially in their decreasing risk order. The concept is similar to the work of Bilder and colleagues on informed Sterrett pooling method with qualitative assays \citep{Bilder2010, McMahan2012}. By testing individuals with a high risk of failure first, mMPA can identify failing individuals more rapidly and efficiently, hence accelerating the deconvolution process.                              

Properties of MP, MPA, and mMPA have been demonstrated and compared using simulations under limited parametric assumptions or clinical implementations/lab testing under specific settings \citep{May2010, Liu2017JAIDS, Holland2019}. Yet the general statistical properties of these methods have not been fully characterized. This paper makes the following contributions to the literature. First, we show that the concept of incorporating additional information into pooled testing can be applied to mMPA for quantitative assays in a similar, but fundamentally different, manner as informed Sterrett retesting for qualitative assays.  Second, we quantify the testing efficiencies of MP, MPA, and mMPA in terms of the average number of assays needed per individual. Third, based on the testing efficiency, we establish the superiority of mMPA to MPA and the bounds on their relative efficiency. Fourth, we establish the difference in the diagnostic accuracy of the three methods compared to individual testing. Finally, we conduct a simulation study to evaluate various factors that affect the testing efficiency of the pooling strategies, such as the prevalence of failure, pool size, assay measurement error, and choice of risk score for mMPA.

The paper is organized as follows. In Section \ref{sec:Notations-and-Definitions}, we present notation and define pool positivity for pooling with quantitative assays. In Section \ref{sec:Method}, we formalize the MP, MPA, and mMPA procedures and the algorithms used for MPA and mMPA. In Section \ref{sec:Test-Efficiencies}, we derive the testing efficiencies for the three testing procedures in terms of average assays needed per individual and focus on the comparisons of testing efficiencies between MPA and mMPA. In Section \ref{sec:Comparison-of-Diagnostic}, we compare the diagnostic accuracy among the three pooling procedures and individual testing. For illustration purposes, we apply these pooled testing strategies to simulated datasets and a real clinical dataset in Section \ref{sec:Illustrations}. We conclude with discussions in Section \ref{sec:Conclusion}.

\section{Notations and definitions} \label{sec:Notations-and-Definitions}

\subsection{Notations \label{subsec:Notations} }

Let $V \in [0, +\infty)$ denote the true result of a quantitative assay with a distribution $\mathbf{F}_{V}$. A failure (or a condition of interest) is defined as $Z = \mathbf{1}(V >C)$ with $Z=1$ indicating a failure and $0$ a non-failure, where $\mathbf{1}(\cdot)$ is the indicator function and $C$ a given cutoff value. In the context of viral load monitoring, for example, we are interested to detect high viral load, so that failure occues when virus exceeds a threshold $C$. The prevalence of failure is $p=1-\mathbf{F}_{V}(C)$. Suppose that $K$ individual samples of equal valume are combined to form a pool. Let $\mathbf{V}= (V_{1}, \dots, V_{K})^\top $ denote the individual test results. In the ideal case of having no measurement errors, let us denote the quantification of the assay applied to the pool as 
\[
    V_{\mathrm{pool}}(K)=\frac{1}{K}\sum_{j=1}^{K}V_{j},
\]
(denoted by $V_{\mathrm{pool}}$ henceforth for simplicity unless confusion arises).

For each individual sample, suppose that a vector of other markers $\mathbf{X}$ is available and correlated with $V$. We assume that a scalar risk score $S$ can be calculated from $\mathbf{X}$ with a known function $g(\cdot)$ and that the resulting $S=g(\mathbf{X})$ is positively associated with $V$, in the sense that for two individuals with $(V_{j},S_{j})$ and $(V_{j'},S_{j'})$,
\begin{equation}
    \mathrm{if}\:S_{j}\ge S_{j'},\:\mathrm{then}\:V_{j}\succeq V_{j'},\label{eq:def.risk.score}
\end{equation}
where ``$V_{j}\succeq V_{j'}$'' denotes the stochastic inequality $\Pr(V_{j}>v)\ge\Pr(V_{j'}>v), \forall v\in [0, \infty)$. If $g(\cdot)$ is unknown, risk estimation methods based on regression \citep[e.g.][]{Friedman2004}, classification \citep[e.g.][]{Breiman1984}, or ensemble learning \citep[e.g.][]{Sinisi2007, Xu2019} can be used to obtain a reasonable functional form of $g(\cdot)$.

For a pool of size $K$, let $\mathbf S=(S_1, \dots, S_K)^\top$ be the risk scores corresponding to $V_1, \dots, V_K$. We use $S_{(1)}, \dots, S_{(K)}$ to denote the (decreasing) order statistics of $S_1, \dots, S_K$, 
    \[ S_{(1)} \ge \dots \ge S_{(K)}. \] 
For each $S_{(j)}$, we denote its concomitant order statistic (corresponding test value) of $V$ by $V_{[j]}$. By the definition \eqref{eq:def.risk.score},
    \[ V_{[1]} \succeq \dots \succeq V_{[K]}.\]

\subsection{Definition of Pool Positivity} \label{subsec:Definition-of-Pos-pool}

To account for the dilution effect, a pool is positive if the pool result $V_\mathrm{pool}$ exceeds
\begin{equation}
    C_{\mathrm{pool}}=C/K.\label{eq:def.viral.fail}
\end{equation}
It is straightforward to verify that when $V_{\mathrm{pool}}\le C_{\mathrm{pool}}$, all individuals in the pool can be classified as non-failures. Otherwise, the pool is positive, prompting the testing of individual samples at the second stage.

The definition (\ref{eq:def.viral.fail}) leads to two differences between pooling strategies with qualitative and quantitative assays. Firstly, as mentioned earlier, a positive pool does not necessarily imply that there is at least one failure in the pool. This can be illustrated using a numerical example. Suppose that $C=$ 1000, $K=5$, and the true results of all individuals are $400$. Then, $V_\mathrm{pool} = 400$ which is greater than $C/5=200$. This pool is positive but contains no failures. The rationale to call this pool positive is that the pool result alone does not allow us to distinguish this pool from other pools that may contain failures (e.g.\ a pool with one sample being 1200 and the other four being 200). Secondly, for pooling with quantitative assays, the pool test result contains information about the possible $number$ of failures, which is different from pooling with qualitative assays wherein the only information that can be learned from a positive pool is that the pool contains \emph{at least} one failure.
%%%%%%%%%%%%%%%%%%%%%%%%%%%%%%%%%%%%%%%%%
\begin{prop} \label{prop:num-assay-bound}
  For pooled testing with quantitative assays, let $K$ be the pool size and $V_{\mathrm{pool}}$ the pool test result. The number of failures in the pool is bounded sharply by
  \begin{equation}
      \left [\mathbf{1}(V_{\mathrm{pool}}>C), ~\lfloor KV_{\mathrm{pool}}/C\rfloor\right ], \label{eq:num-assays-bounds}
    \end{equation}
  where $\lfloor\cdot\rfloor$ is the floor operation.
\end{prop}
%%%%%%%%%%%%%%%%%%%%%%%%%%%%%%%%%%%%%%%%%
The bounds are the result of maximizing  $\sum_{j=1}^{K}\mathbf{1}(V_{j}>C)$ (or $\sum_{j=1}^{K}\mathbf{1}(V_{j}\le C)$) subject to the constraints of $\sum_{j=1}^{K}V_{j}=KV_{\mathrm{pool}}$ and $V_{j}\ge 0$, for $j=1,\dots,K$. 
%see Appendix \ref{sec:Proof-num-assay-bounds}. 
The lower bound suggests that for a pool with $C/K\le V_{\mathrm{pool}}\le C$, the pool is called positive but possibly contains no failures. The upper bound suggests that with a pool test result $V_{\mathrm{pool}}>C$, the maximum possible number of failures is $\lfloor KV_{\mathrm{pool}}/C\rfloor$. Hence, after a positive pool, testing individual samples sequentially with a stopping algorithm can save cost when it comes to test assays.

\section{Formulation of Pooling Strategies MP, MPA and mMPA}\label{sec:Method}

\begin{figure}[ht!]
\begin{centering}
    \includegraphics[width=7in]{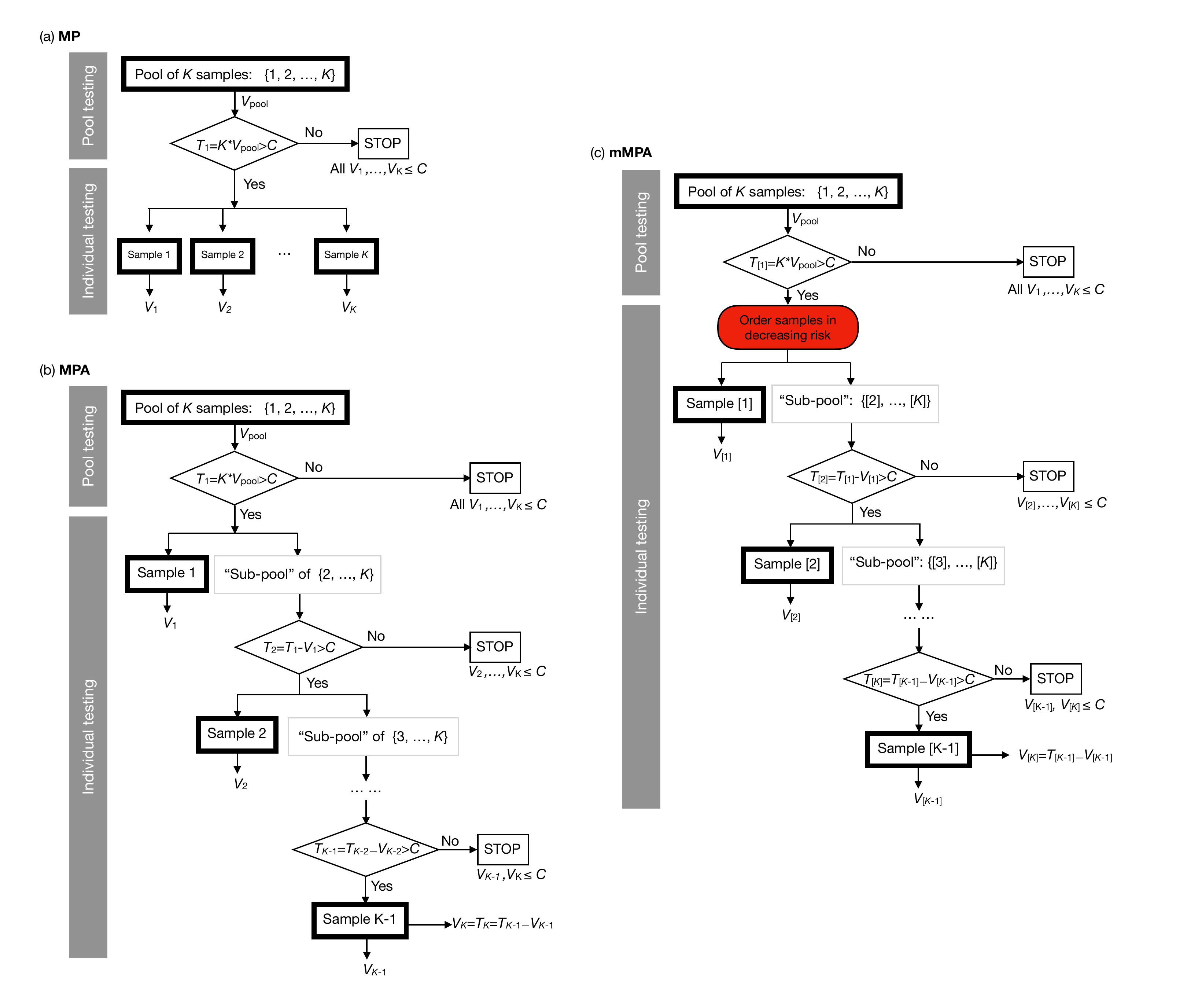}
\par\end{centering}
\centering{}\caption{Schematic diagrams of (a) MP, (b) MPA and (c) mMPA. Boxes with bold border lining indicate that a test is conducted on a sample or a pool of samples. Boxes with light border lining indicate a hypothetical sub-pool of remaining samples. $T_j=\sum_{l=j}^K V_l$ and $T_{[j]}=\sum_{l=j}^K V_{[l]}$, where $V_j$ denotes the test result of $j$th individual (``Sample $j$'') in the pool and $V_{[j]}$ the result of  individual sample with the $j$th highest risk (``Sample $[j]$'').   \label{fig:mMPA-procedure}}
\end{figure}

MP, MPA, and mMPA differ in the second stage after a pool tests positive. Of the three methods, MP is the most straightforward to implement. After a pool tests positive, all individual samples are tested at the second stage to identify individual failures; see Figure~\ref{fig:mMPA-procedure}(a). For a positive pool, the total number of assays needed is thus $K+1$; otherwise, only one assay is needed.

MPA is similar to MP, but after a pool tests positive, individual samples are tested sequentially with a deconvolution and stopping algorithm.  Figure~\ref{fig:mMPA-procedure}(b) provides a schematic description of the MPA procedure. To formulate the algorithm, let us denote $T_{j}=\sum_{l=j}^{K}V_{l} = KV_{\mathrm{pool}}$. When $V_\mathrm{pool}>C_\mathrm{pool}$ or equivalently $T_{1}>C$, the pool is positive and we proceed to test individual samples at the second stage.  After obtaining $V_{1}$ from the first individual sample, we calculate $T_{2}=T_{1}-V_{1}$. Note that we can regard the remaining samples as a ``sub-pool'' with a pool size of ($K-1$), on which if we hypothetically performed a pooled test, we would obtain a result of $T_{2}/(K-1)$. This ``sub-pool'' is negative if $T_2 \le C$;  otherwise, the ``sub-pool'' is positive and we continue individual testings. After obtaining the result $V_{2}$ from the second individual sample, we calculate $T_{3}=T_{2}-V_{2}$, and so on. We stop after testing the $j$th sample when $T_{j+1}\le C$ and conclude all the remaining samples are negative.  

\subsection{mMPA Procedure} \label{subsec:mMPA-Procedure}

The pooling strategy of mMPA improves upon MPA by incorporating other available markers $\mathbf{X}$ into the test procedure; see Figure~\ref{fig:mMPA-procedure}(b) for a schematic description of mMPA procedure. When a pool tests positive, mMPA orders the individual samples in the decreasing rank order of their estimated risks (calculated from $\mathbf X_1, \dots, \mathbf X_K$ with a function $g(\cdot)$) and tests the individual samples in that order. Similarly, let us denote $T_{[j]} = \sum_{l=j}^{K} V_{[l]}$ with $T_{[1]}=T_{1}$. When $T_{[1]}>C$, the pool is positive, prompting the testing of individual samples. After obtaining $V_{[1]}$ from the individual with the highest estimated risk, we calculated $T_{[2]} = T_{[1]}-V_{[1]}$. We stop after testing the sample that has the $j$th highest risk score when $T_{[j+1]} = T_{[j]}-V_{[j]} \le C$ and conclude all the remaining samples with risk scores $<S_{(j)}$ as negative. 

The MPA and mMPA procedures are similar to the informed Sterrett pooling procedures for qualitative assays \citep{Bilder2010}. However, due to the quantitative nature of the assays, we are able to calculate $T_{j+1}=T_{j}-V_{j}$, or $T_{[j+1]}=T_{[j]}-V_{[j]}$, avoiding the need to de facto form and test the ``sub-pools'' of remaining samples.

\section{Test Efficiencies of MP, MPA, mMPA} \label{sec:Test-Efficiencies}

For a given pool size of $K$, let $\phi(K)$ denote the testing efficiency of a pooling strategy, defined as the average tests needed per individual. For individual testing without pooling, $\phi_\mathrm{IND} \equiv 1$. For comparing different testing strategies, a smaller value of $\phi$ means a higher testing efficiency and $(1-\phi)$ is the cost saving in testing assays compared with individual testing. 

\subsection{MP procedure}\label{subsec:MP-Procedure}

For MP, one can show that the average number of assays needed per individual is
\[
   \phi_\mathrm{MP}(K) = \frac{1}{K} + \Pr(T_{1}>C).
\]
To obtain an estimate of $\phi_\mathrm{MP}$, suppose that we have a dataset of $N$ random individuals with $N\gg K$. To distinguish from the test results $(V_1, \dots V_K)$ of $K$ sample in a pool, we use $Y_i$ to denote the test result of the $i$th individual in the dataset for $i=1, \dots, N$. We can first estimate $\mathbf{F}_{T_{K}}(C)$ empirically by $\widehat{\mathbf{F}}_{T_{K}}(C) = \widehat{\mathbf{F}}_{V}(C) = \frac{1}{N}\sum_{i=1}^{N}\mathbf{1}(Y_{i} \le C)$ and then $\mathbf{F}_{T_{j}}(C)$ sequentially by
\begin{equation}
\mathbf{\widehat{F}}_{T_{j}}(C)=(\mathbf{\widehat{F}}_{T_{(j+1)}}*\mathbf{\widehat{F}}_{V})(C)=\frac{1}{N}\sum_{i=1}^{N}\widehat{\mathbf{F}}_{T_{(j+1)}}(C-Y_{i})\mathbf{1}(Y_{i}\le C).\label{eq:emp.distn.Tj}
\end{equation}
Then the average number of assays needed per individual using MP for a pool of size $K$ is estimated by
\[
    \widehat \phi_\mathrm{MP}(K) = \frac{1+K}{K} - \mathbf{\widehat{F}}_{T_{1}}(C).
\]
When most of the pools are negative (i.e.\ $\Pr(T_1\le C) \uparrow 1$), MP can be highly efficient because $\phi_\mathrm{MP}(K) \downarrow 1/K$. On the other hand, when more than $(K-1)/K$ of the pools are positive, individual testing without pooling is more efficient than MP. 

%[Random thought: If we treat $K$ as continuous, can we find the optimal pool size by solving $\frac{\partial \phi(K)}{\partial K} =0$?]

\subsection{MPA procedure}\label{subsec:MPA-Procedure}

MPA tests individual samples sequentially for a positive pool; see Figure \ref{fig:mMPA-procedure}(b). After testing the $j$th sample and obtaining $V_j$, we calculate $T_{j+1} = T_j-V_j$ with $T_1=K V_{\mathrm{pool}}$. If $T_{j+1} > C$, we continue to test the $(j+1)$th sample; otherwise, we stop individual testing. That implies that the total number of assays needed for a pool is $1 + \mathbf{1}(T_1 > C) + \dots + \mathbf{1}(T_{K-1} > C)$. Per individual, the average number of assays needed is 
\[
    \phi_{\mathrm{MPA}}(K) = 1-\frac{1}{K}\sum_{j=1}^{K-1}\mathbf{F}_{T_{j}}(C).
\]
Given test results data of $N$ individuals, we can substitute $\mathbf{F}_{T_{j}}(C)$ by \eqref{eq:emp.distn.Tj} and estimate $\phi_{\mathrm{MPA}}(K)$ by
\[
    \widehat\phi_{\mathrm{MPA}}(K) = 1 - \frac{1}{K}\sum_{j=1}^{K-1} \mathbf{\widehat{F}}_{T_{j}}(C).
\]
Furthermore, because $T_1\ge \dots \ge T_K$, we have $\mathbf{F}_{T_1}(C) \le \dots \le \mathbf{F}_{T_K}(C).$ It is straightforward to verify (see Appendix \ref{subsec:phi.mpa}) that MPA is more efficient than both MP and individual testing, 
\begin{equation}\label{eq:phi.MPA}
 \phi_\mathrm{MPA}(K) \le \min (\phi_\mathrm{PM}(K), \phi_\mathrm{IND}). 
\end{equation}

\subsection{mMPA procedure}

When a pool tests positive, instead of testing the individual samples in a random order, mMPA tests the samples in the decreasing order of their estimated risks. Similar to MPA, for a given pool of size $K$, the total number of assays needed for a pool is $1 + \sum_{j=1}^{K-1}\mathbf{1}(T_{[j]}>C) $.  That implies that the testing efficiency of mMPA is
\begin{equation*} 
    \phi_\mathrm{mMPA}(K)=1-\frac{1}{K}\sum_{j=1}^{K-1}\Pr(T_{[j]} \le C). 
\end{equation*}
Because the ordering of individual samples depends on the relative magnitudes of the risk scores, $V_{[1]},\dots,V_{[K]}$ are no longer independently nor identically distributed. To calculate $\phi_{\mathrm{mMPA}}$, we further introduce two distribution functions, $\mathbf{F}_{V|s}(v)=\Pr(V \le v\mid S=s)$ which is the conditional distribution of $V$ among those with risk score $S=s$ and $\mathbf{F}_{V:s}(v)=\Pr(V \le v\mid S<s)$ which is the ``truncated'' distribution of $V$ among those with risks of $S<s$. Further, we denote the risk score distribution by $\mathbf{F}_{S}$. With some algebra (see Appendix \ref{subsec:Derivation-of-exp-mmpa}), we show that for a given pool size $K$,
\begin{equation}
	%\Pr(T_{[j]} \le C) =
	\phi_\mathrm{mMPA}(K)=1-\frac{1}{K}\sum_{j=1}^{K-1}\int_{\mathbf{F}_{S}(s)=0}^{1}(A_{j}^{s}*\mathbf{F}_{V|s})(C)\mathrm{d}\mathcal{B}_{j,(K+1-j)}\{\mathbf{F}_{S}(s)\},\label{eq:expc-mmpa}
\end{equation}
where $\mathcal{B}_{j,(K+1-j)}(\cdot)$ is the Beta distribution with parameters~$j$ and $(K+1-j)$, and
$
    A_{j}^{s}(v)=\Pr(T_{[j+1]} \le v|S_{(j)}=s)
$
for $j=1,\dots,(K-1)$ with $A_{K}^{s}(v)=\mathbf{F}_{V:s}(v)$. 

Suppose that we have data from $N$ individuals, from whom we obtain both their test results and corresponding risk scores $(Y_{i},S_i)$, $i=1,\dots,N$. We can estimate $A_{K}^{s}(v)$ and ${A}_{j}^{s}(v)$, respectively, by 
\begin{align*}
	\widehat{A}_{K}^{s}(v) &=\mathbf{\widehat{F}}_{V:s}(v)=\frac{\sum_{i=1}^{n}\mathbf{1}(Y_{i} \le v,S_i<s)}{\sum_{i=1}^{n}\mathbf{1}(S_i<s)}, \\ 
    \widehat{A}_{j}^{s}(v) &=\frac{1}{\sum_{i=1}^{n}\mathbf{1}(S_i<s)}\sum_{i=1}^{n}\widehat{A}_{j+1}^{s}(v-Y_{i})\mathbf{1}(Y_{i} \le v,S_i<s),\quad j=1,\dots,(K+1).
\end{align*} We can then empirically estimate the conditional distribution function $\mathbf{F}_{V|s}(v)$ by $\mathbf{\widehat{F}}_{V|s}(v)=\sum_{i=1}^{n}\mathbf{1}(Y_{i}\le v,S_i=s)/\sum_{i=1}^{n}\mathbf{1}(S_i=s)=\mathbf{1}(Y_{i}\le v,S_i=s)$. Using the result (\ref{eq:appdx.PrTj}) in Appendix \ref{subsec:Derivation-of-exp-mmpa} and plugging $\widehat{A}_{j}^{s}(\cdot)$ and $\mathbf{\widehat{F}}_{V|s}(\cdot)$ in  (\ref{eq:expc-mmpa}) , we obtain an estimate of  $\Pr(T_{[j]}\le C)$ as
\[
\mathbf{\widehat{F}}_{T_{[j]}}(C) = \sum_{i=1}^{N}\frac{S_i^{j-1}(1-S_i)^{K-j-1}}{b(j,K+1-j)}\mathbf{\widehat{F}}_{T_{[j+1]}\mid S_{(j)}=s}(C-Y_{i}),
\]
where $b(\cdot,\cdot)$ is the beta function. Further, we obtain
\[
\widehat{\phi}_{\mathrm{mMPA}} = 1-\frac{1}{K}\sum_{j=1}^{K-1}\mathbf{\widehat{F}}_{T_{[j]}}(C), 
\]
where $\frac{1}{K}\sum_{j=1}^{K-1}\mathbf{\widehat{F}}_{T_{[j]}}(C)$ is the relative reduction in the number of assays needed by mMPA compared with individual testing without pooling.

\subsection{mMPA with Different Risk Scores}

The mMPA procedure does not rely on a specific risk score. However, different choices of risk scores lead to different orders of individual samples for  sequential testing and result in different testing efficiencies.

Suppose that there are two risk scores $S^{(a)}$ and $S^{(b)}$ that can be derived from available low-cost markers. Let us denote the data for individual samples in a pool by triplets of $(V_{j},S_{j}^{(a)}, S_{j}^{(b)})$, $j=1,\dots,K$, and the concomitant ordered $V$ values by $V_{[1]}^{(a)},\dots,V_{[K]}^{(a)}$ and $V_{[1]}^{(b)},\dots,V_{[K]}^{(b)}$ using scores $S^{(a)}$ and $S^{(b)}$, respectively.
\begin{defn}
\label{def:stronger score} For two risk scores $S^{(a)}$ and $S^{(b)}$ both satisfying the definition $\eqref{eq:def.risk.score}$, we define $S^{(a)}$ to be a \emph{stronger} risk score than $S^{(b)}$ if 
\begin{equation}
\sum_{j=1}^{k}V_{[j]}^{(a)}\succeq\sum_{j=1}^{k}V_{[j]}^{(b)},\quad\forall k=1,\cdots,K.\label{eq:def.stronger.score}
\end{equation}
Because of the constraint $\sum_{j=1}^{K}V_{[j]}^{(a)}=\sum_{j=1}^{K}V_{[j]}^{(b)}=\sum_{j=1}^{K}V_{j}$, $\eqref{eq:def.stronger.score}$ also implies that $\sum_{j=k}^{K}V_{[j]}^{(a)}\preceq\sum_{j=k}^{K}V_{[j]}^{(b)}$, $ k=1,\cdots,K$. This definition means that a stronger risk score is better at separating those individuals with greater values of $V$'s from those with smaller values.
\end{defn}

\begin{thm}
\label{thm:mMPA-for-two-risk-scores}For two risk scores $S^{(a)}$ and $S^{(b)}$, if $S^{(a)}$ is a stronger risk score than $S^{(b)}$, then for the same pool size $K$, mMPA based on $S^{(a)}$ is more efficient than based on $S^{(b)}$,
\begin{equation}
\phi_{\mathrm{mMPA}}^{(a)} \le \phi_{\mathrm{mMPA}}^{(b)}, \label{eq:cor1.2.risk.scores}
\end{equation}
where $\phi_{\mathrm{mMPA}}^{(a)}$ and $\phi_{\mathrm{mMPA}}^{(b)}$ are the average numbers of assays needed per individual by mMPA based on $S^{(a)}$ and $S^{(b)}$, respectively.
\end{thm}
A brief proof is provided in Appendix \ref{sec:Proof-of-Theorem-2}. The following property about the testing efficiency of mMPA follows immediately.
\begin{cor}
\label{cor:E-M-bounds} For a given pool size $K$, the testing efficiency of mMPA is bounded by
\begin{equation}
   \frac{ 1+Kp-p^{K} }{K} \le \phi_{\mathrm{mMPA}} \le 1-\frac{1}{K} \sum_{j=1}^{K-1}\mathbf{F}_{T_{j}}(C). \label{eq:cor2}
\end{equation}
\end{cor}

A proof is provided in Appendix \ref{sec:Proof-of-Corollary-bound}.

\subsection{Efficiency Comparison between mMPA and MPA}

MPA can be regarded as a special case of mMPA when the risk scores $\mathbf{S}$ being used are independent of $\mathbf{V}$, so that ordering the samples by $\mathbf{S}$ leads to $V_{[j]} \stackrel d = V$ and $T_{[j]} \stackrel d = T_j$ for $j=1, \dots, K$. In this case, $\mathbf{S}$ is the weakest risk score among all risk scores that satisfy the definition~$(\ref{eq:def.risk.score})$. The following result follows immediately.
\begin{cor}
\label{cor:mMPA-more-efficient-than-MPA} Using the same pool size $K$, mMPA is more efficient than MPA, which is more efficient than MP and individual testing, 
\begin{equation}
\phi_\mathrm{mMPA} \le \phi_\mathrm{MPA} \le (\phi_\mathrm{MP} \land \phi_\mathrm{IND}).
\label{eq:cor.mmpa-better-than-mpa}
\end{equation}
\end{cor}

It is not difficult to verify that when measurement errors exist in testing assays, the relative efficiency~(\ref{eq:cor.mmpa-better-than-mpa}) still holds, although the measurement errors can lead to misclassification errors. 

\section{Comparison of Diagnostic Accuracy between mMPA and MPA\label{sec:Comparison-of-Diagnostic}}

In practice, diagnostic testing based on quantitative assays is subject to measurement error, which leads to misclassifications. Let us denote an assay's measurement error by $\varepsilon$. In this paper, we assume that $\varepsilon$ is multiplicative and symmetric around one in the sense that $\varepsilon \stackrel d= \varepsilon^{-1}$ (e.g.~$\varepsilon\sim$ lognormal distribution). The observed test result is $\widetilde V = V\varepsilon$. Following convention, the sensitivity (SENS), specificity (SPEC), positive (PPV) and negative predictive values (NPV) of individual testing are defined as
\begin{align*}
    &\mathrm{SENS}_\mathrm{IND} = \Pr(\widetilde V>C|V>C), \quad 
    \mathrm{SPEC}_\mathrm{IND} = \Pr(\widetilde V\le C|V\le C), \\
    &\mathrm{PPV}_\mathrm{IND} =\Pr(V>C | \widetilde V >C), \quad 
    \mathrm{and} \quad 
    \mathrm{NPV}_\mathrm{IND} =\Pr(V\le C| \widetilde V\le C).
\end{align*}

For pooled testing, we define diagnostic accuracy measures from the perspective of an individual or their treating physician. Suppose a physician prescribes a test for a patient, knowing that the lab performs a pooled testing but without knowing the pool test result or the results of other patients in the pool. Given only the final test result of the patient, the diagnostic accuracy measures that relate the patient's test result from pooled testing and its true status are defined as 
\begin{align*}
    \mathrm{SENS}_{\mathrm{pool}} & =\Pr(\textrm{A patient tests positive through pooled testing} \mid \textrm{patient's true status } V>C),\\
    %& =\Pr(\textrm{pool tests positive }and\textrm{ the sample tests positive}|V>C),\\
    \mathrm{SPEC}_{\mathrm{pool}} & =\Pr(\textrm{A patient tests negative through pooled testing} \mid \textrm{patient's true status } V\le C),\\
    %& =\Pr(\textrm{pool tests negative }or\textrm{ (pool tests positive but the sample tests negative)}|V\le C),\\
    \mathrm{PPV}_{\mathrm{pool}} & =\Pr(\textrm{A patient's } V>C \mid \textrm{pooled testing classifies the patient as positive}),\\
    \mathrm{NPV}_{\mathrm{pool}} & = \Pr(\textrm{A patient's } V\le C \mid \textrm{pooled testing classifies the patient as negative}).
\end{align*}
With these definitions, we establish the diagnostic accuracy of individual testing, MP, MPA, and mMPA. 
\begin{thm}
    With the same pool size for MP, MPA and mMPA, the diagnostic accuracy of IND (individual testing), MP, MPA, and mMPA are ordered as follows: \\
    For sensitivity,
    \[ \mathrm{IND}\ge \mathrm{MP} \ge \{\mathrm{mMPA}, \mathrm{MPA} \}; \]
    for specificity and PPV,
    \[ \{\mathrm{mMPA},\mathrm{MPA} \} \ge \mathrm{MP} \ge \mathrm{IND}; \]
    and for NPV if we can assume that the condition \eqref{eq:NPR.assumption} in Appendix \ref{sec:diagnosis.NPV} holds, 
    \[ \mathrm{IND}\ge \mathrm{MP} \ge \{\mathrm{mMPA}, \mathrm{MPA} \}, \]
    where ``$\ge $'' denotes ``superior or equivalent to''. \label{thm:The-diagnostic-accuracies}
\end{thm}
A proof is provided in Appendix \ref{appndx:diagnostic}. It should be pointed out that although a comparison of diagnostic accuracy between MPA and mMPA is not established, mMPA can outperform MPA with a properly chosen risk score while requiring fewer assays. This property has been illustrated through Monte Carlo simulations in \citet{Liu2017JAIDS} under specific parametric simulation setups. It is also worthwhile to remark that theoretical results of pooled testing with qualitative assays may not apply to pooling with quantitative assays. For example, it has been shown that for MP with qualitative assays, $\mathrm{SENS}_\mathrm{MP} = \mathrm{SENS}_\mathrm{IND}^2$ \citep{kim2007comp, Johnson1991, McMahan2012}. However, this result does not hold for pooled testing with quantitative assays. As we show in Appendix \ref{appndx:sens}, the sensitivity of MP with quantitative assay is bounded by $ \mathrm{SENS}_\mathrm{IND}^2 \le \mathrm{SENS}_\mathrm{MP}\le \mathrm{SENS}_\mathrm{IND}$.

\section{Illustrations \label{sec:Illustrations} }

\subsection{Numerical Examples \label{sec:Simulation-Study}}

We conduct four simulation studies to illustrate the properties of MP, MPA, and mMPA. We simulate test results of a quantitative assay using an exponential distribution with a scale parameter $\theta$.  We choose exponential distribution so that for comparison, we can analytically derive the properties of pooled testing using the fact that the convolution of independent exponential distributions with a common rate parameter leads to a gamma distribution.

\begin{figure}
\begin{centering}
\includegraphics[width=6.5in]{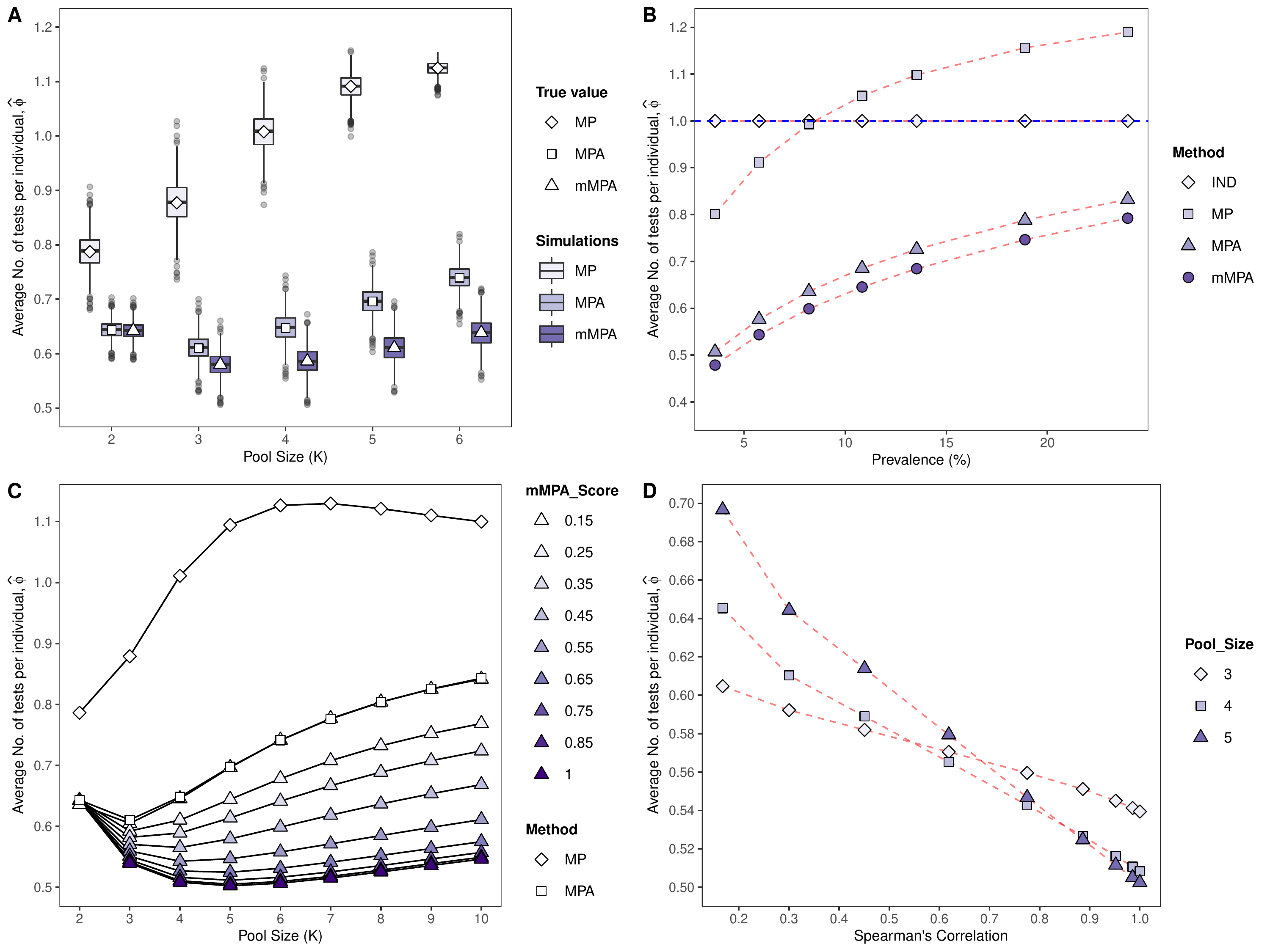}
\par\end{centering}
\caption{(A) Testing efficiency comparison of MP, MPA and mMPA for pool sizes ranging from 2 to 6. (B) Impact of failure prevalence on IND, MP, MPA and mMPA, where a pool size of 4 is used for the three pooling methods. (C) Comparison of mMPA with different risk scores. The numbers in the legend show the weights used to simulate the corresponding risk scores. For reference, the testing efficiencies of MP and MPA are also shown. (D) Impact of the strength of risk scores (Spearman's correlation) on mMPA for pool sizes of 3, 4, and 5. \label{fig:simulationf}}
\end{figure}

\emph{Simulation Study I}:  We use exponential distribution with $\theta=400$ to simulate test results and a cutoff of $C=1,000$ to define a disease/failure. This leads to a failure prevalence of $1-\mathbf{F}_{V}(C) \approx 8\%$. We simulate observations of $(Y, S)$ from 300 individuals, where $S = \lambda\times \mathrm{Rank}(Y)/300 + (1-\lambda)U$, $U$ has a uniform(0, 1) distribution, and $\lambda\in[0,1]$ controls the correlation between $Y$ and $S$.  For Simulation Study~I, we choose $\lambda=0.25$ which leads to a moderate Spearman's correlation of about $0.30$ between $Y$ and $S$. We consider five pool sizes ($K$ ranging from 2 to 6) to group the 300 individuals for pooled testing. For each pool size, we estimate the average number of assays needed per individual ($\widehat \phi$'s) using the formulae in Section 4.1-4.3 for MP, MPA, and mMPA. We repeat this for the 2,000 simulations. Figure \ref{fig:simulationf}(A) shows the distributions of the calculated $\widehat \phi$'s using box-plots for MP, MPA, and mMPA. For reference, the true values of $\phi$'s are calculated analytically for each pooling method and added to the Figure. Figure \ref{fig:simulationf}(A) shows that with properly chosen pool sizes, the pooling strategies MP, MPA, and mMPA can effectively reduce the number of assays needed to ascertain all individuals' status. For this specific simulation setup, the optimal pool sizes are 2, 3 and 3 for MP, MPA, and mMPA, respectively, resulting in reduction in cost of assays by 21\%, 39\%, and 42\% compared with individual testing (or equivalently increasing the testing capacity by testing 127\%, 164\% and 172\% individuals with the same number of assays as individual testing). MP may be less efficient than individual testing when a large pool size is used and most pools are positive. 

\emph{Simulation Study II}: We evaluate the impact of failure prevalence on the testing efficiency of individual testing, MP, MPA, and mMPA. We choose $\lambda=0.25$ and a pool size of $K=4$. We simulate test results using exponential distributions with $\theta \in \{ 300, 350, 400, 450, 500, 600, 700 \}$, which result in failure prevalences of 3.6\%, 5.7\%, 8.2\%, 11\%, 14\%, 19\% and 24\%, respectively. To ease the comparisons of testing efficiency with various failure prevalences, we choose a large sample size (so as to have a small estimation variation) and simulate 2,000 pairs of $(Y,S)$ for each value of $\theta$. The results are shown in Figure \ref{fig:simulationf}(B). As expected, all pooling methods can be highly efficient when the failure prevalence is low. As the prevalence increases, their efficiencies decrease. MP can be less efficient then individual testing when the prevalence is high. MPA and mMPA are more efficient than both individual testing and MP. Similar pattern of the impact of failure prevalence on testing efficiency is observed for other pool sizes. 

\emph{Simulation Study III}: We evaluate the impact of risk scores being used by mMPA on its testing efficiency. Risk scores of different strengths are simulated using the formula as Simulation Study I with $\lambda \in \{0.15,0.25,\dots,0.85,1\}$.  The resulting Spearman's correlations between the simulated $Y$ and $S$ are $\{0.16,0.30,0.45, 0.62,0.77,0.89, 0.95,0.99, 1\}$, respectively. Again to ease the comparisons of mMPA with different risk scores, we choose a large sample size of 2,000 for each value of $\lambda$. Figure \ref{fig:simulationf} (C) shows $\widehat \phi_\mathrm{mMPA}$ with risk scores of different strengths. For reference, $\widehat\phi_\mathrm{MP}$ and $\widehat\phi_\mathrm{MPA}$ are also shown in the Figure. Clearly, stronger risk scores lead to a higher testing efficiency of mMPA. With a stronger risk score, a larger pool size also could possibly be used to improve testing efficiency. Figure \ref{fig:simulationf} (D) shows the testing efficiency of mMPA versus the Spearman's correlations of $(Y, S)$ for pool sizes of 3, 4, and 5. The Figure shows that the choice of risk score impacts the testing efficiency of mMPA with Spearman's correlation being a reasonable metric for evaluating risk score's strength. 

\begin{figure}
\begin{centering}
\includegraphics[width=6.3in]{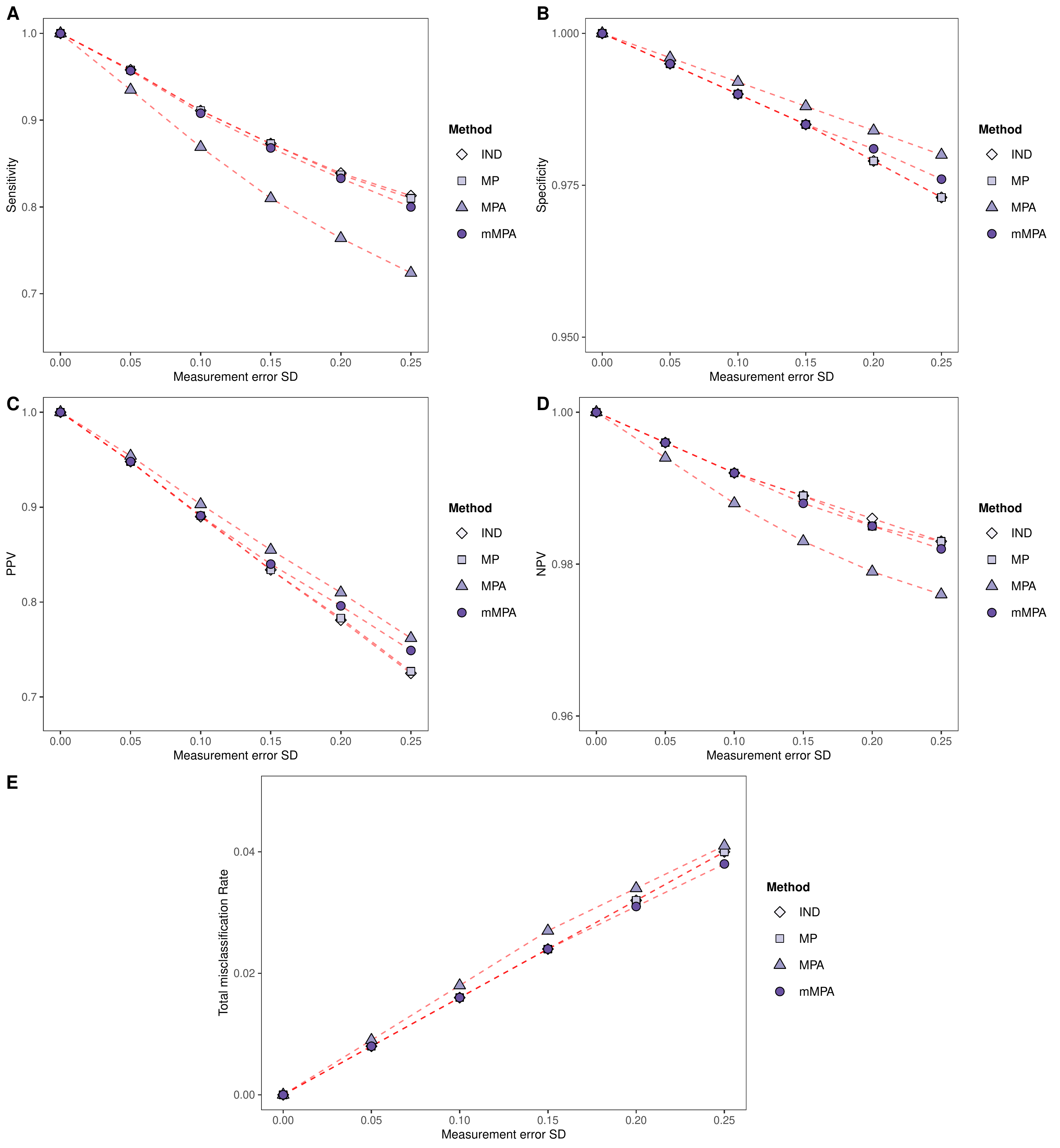}

\par\end{centering}
\caption{Impact of measurement error on classification accuracy: (A) sensitivity, (B) specificity, (C) PPV, (D) NPV, and (E) total misclassification rate. Measurement errors are simulated using log-normal distributions with standard deviations (in log scale) ranging from 0 to 0.25. A pool size of 5 is used for the pooling methods (MP, MPA, and mMPA).   \label{fig:simulationErr}}
\end{figure}

\emph{Simulation Study IV}: We evaluate the impact of measurement error on classification accuracy by simulating the errors on individual and pool samples using log-normal distribution. We choose $\theta=400$ and $K=5$. The measured test results of individual and pool samples are generated by 
    $$ \tilde V = V\varepsilon, $$
where $\log\varepsilon \sim \mathrm{N}(0, \sigma^2)$. We use $\sigma \in \{0, 0.05, 0.1, 0.15, 0.2, 0.25\}$ to simulate different magnitudes of measurement errors. The results of classification accuracy are shown in Figure \ref{fig:simulationErr}.  Overall, the sensitivity (Figure \ref{fig:simulationErr}(A)) and NPV (Figure \ref{fig:simulationErr}(D)) of IND and MP are higher than those of MPA and mMPA. MPA and mMPA have higher specificity (Figure \ref{fig:simulationErr}(B)) and PPV (Figure \ref{fig:simulationErr}(C)) than IND and MP. These findings are consistent with our results in Section 5. Moreover, in terms of total misclassification rate, mMPA is better than the other methods, although the differences are small (Figure \ref{fig:simulationErr}(E)). 

\subsection{Application to HIV Treatment Monitoring}\label{sec:Empirical-Application}

We apply the pooling methods to a HIV dataset from the Miriam Hospital Immunology Clinic in Rhode Island, United States \citep{Liu2013} to estimate the number of HIV VL assays that would be required had pooling strategies been used for HIV treatment monitoring. The data were obtained from 597 patients who had been on HIV treatment for at least 6 months and had CD4 count, CD4\% at their most recent clinic visit, CD4 count, CD4\% from six months before, and HIV VL measurements available. A summary of the patients' characteristics can be found in \citet[Table 2]{Liu2013}. We use the WHO guidelines of VL greater than $C=1,000$ copies/mL to define ART failure, which is relevant for resource limited settings where VL pooling might be more needed. For this population, the estimated prevalence of treatment failure is about 21\%.

Using this dataset, we estimate the average number of assays that would be needed when using MP, MPA and mMPA. For mMPA, two risk scores from the data are considered: (a) the negative value of CD4 count and (b) a composite risk score calculated using the following formula (which was derived from fitting a logistic model of ART failure on CD4 count, CD4\% and their 6-month (`current' minus `6 months ago') changes; see Liu et al., 2013, Sec 6.)
$$-0.0021\times [\textrm{CD4}] -0.049\times [\textrm{CD4 \%}]-0.055\times [\textrm{6-mo CD4 change}]-1.4\times [\textrm{6-mo CD4\% change}].$$  The two risk scores have Spearman's correlations with HIV VL of 0.266 and 0.302, respectively. Moreover for mMPA, we consider using (c) an oracle score, which is generated using the ranks of VL values. 

Table \ref{tab:Miriam-Data-mMPA} shows the testing efficiencies of mMPA using the risk scores (a), (b), and (c) for different pool sizes. The bootstrap method with 1,000 resamples is used to calculate the 95\% confidence intervals (95\% CIs). For mMPA using the composite risk score (b), the calculated values of $\widehat \phi_\mathrm{mMPA}$ are slightly lower (more efficient) than the corresponding $\widehat \phi_\mathrm{mMPA}$ of the same pool size using the risk score (a) (CD4 alone). The optimal pool sizes are 5 for both risk scores. The resulting testing efficiencies are $0.563$ (95\% CI = $0.433, 0.671$) and 0.549 ($0.428, 0.668$) assay per individual, respectively, with an insignificant difference of 0.014 $(-0.028, 0.054)$. That means, with a properly chosen risk score and pool size, mMPA could reduce the cost of VL assays by up to 45\% compared with individual testing. Or, using the same number of assays, about twice as many people could be monitored using mMPA. For this data, using the composite score (b) does not significantly improve the testing efficiency than the simple score of CD4 count alone, which is much easier for clinical implementation. The testing efficiency using the oracle risk score (c) for a pool size of 5 is estimated to be 0.415 (0.336, 0.495) assay per individual, which shows the maximum possible reduction of VL assays needed using mMPA is about $60\%$.

\begin{table}
\caption{Miriam Data: Pooled HIV VL testing using mMPA with (a) CD4 count as a risk score; (b) a composite risk score derived from CD4 count, CD4\%, and their 6-month changes; and (c) an oracle score that perfectly predicts VL rank orders.  
\label{tab:Miriam-Data-mMPA}}
\centering{}%
\begin{tabular}{llccc}
\hline
Risk Score &  Spearman's correlation  & Pool Size & $\widehat \phi$ & 95\% C.I.\tabularnewline
\hline
(a) (negative) CD4 & 0.27 & 3 & 0.598 & (0.502, 0.681) \tabularnewline
 &  & 4 & 0.569 & (0.456, 0.672)\tabularnewline
 &  & 5 & 0.563 & (0.433, 0.671)\tabularnewline
 &  & 6 & 0.567 & (0.423, 0.687)\tabularnewline
 &  & 7 & 0.576 & (0.423, 0.699)\tabularnewline
 &  & 8 & 0.587 & (0.426, 0.714)\tabularnewline
\hline
(b) Composite Score & 0.30 & 3 & 0.593 & (0.502, 0.680)\tabularnewline
 &  & 4 & 0.559 & (0.452, 0.664)\tabularnewline
 &  & 5 & 0.549 & (0.428, 0.668)\tabularnewline
 &  & 6 & 0.551 & (0.417, 0.678)\tabularnewline
 &  & 7 & 0.557 & (0.414, 0.693)\tabularnewline
 &  & 8 & 0.567 & (0.416, 0.705)\tabularnewline
\hline
% &  &  & (Lower bound) & \tabularnewline
(c) Oracle Score  & 1 & 3 & 0.541 & (0.463, 0.612)\tabularnewline
 &  & 4 & 0.462 & (0.384, 0.541)\tabularnewline
 &  & 5 & 0.415 & (0.336, 0.495)\tabularnewline
 &  & 6 & 0.383 & (0.304, 0.464)\tabularnewline
 &  & 7 & 0.362 & (0.282, 0.443)\tabularnewline
 &  & 8 & 0.345 & (0.266, 0.427)\tabularnewline
\hline
\end{tabular}
\end{table}

\begin{figure}
\begin{centering}
\includegraphics[width=4.5in]{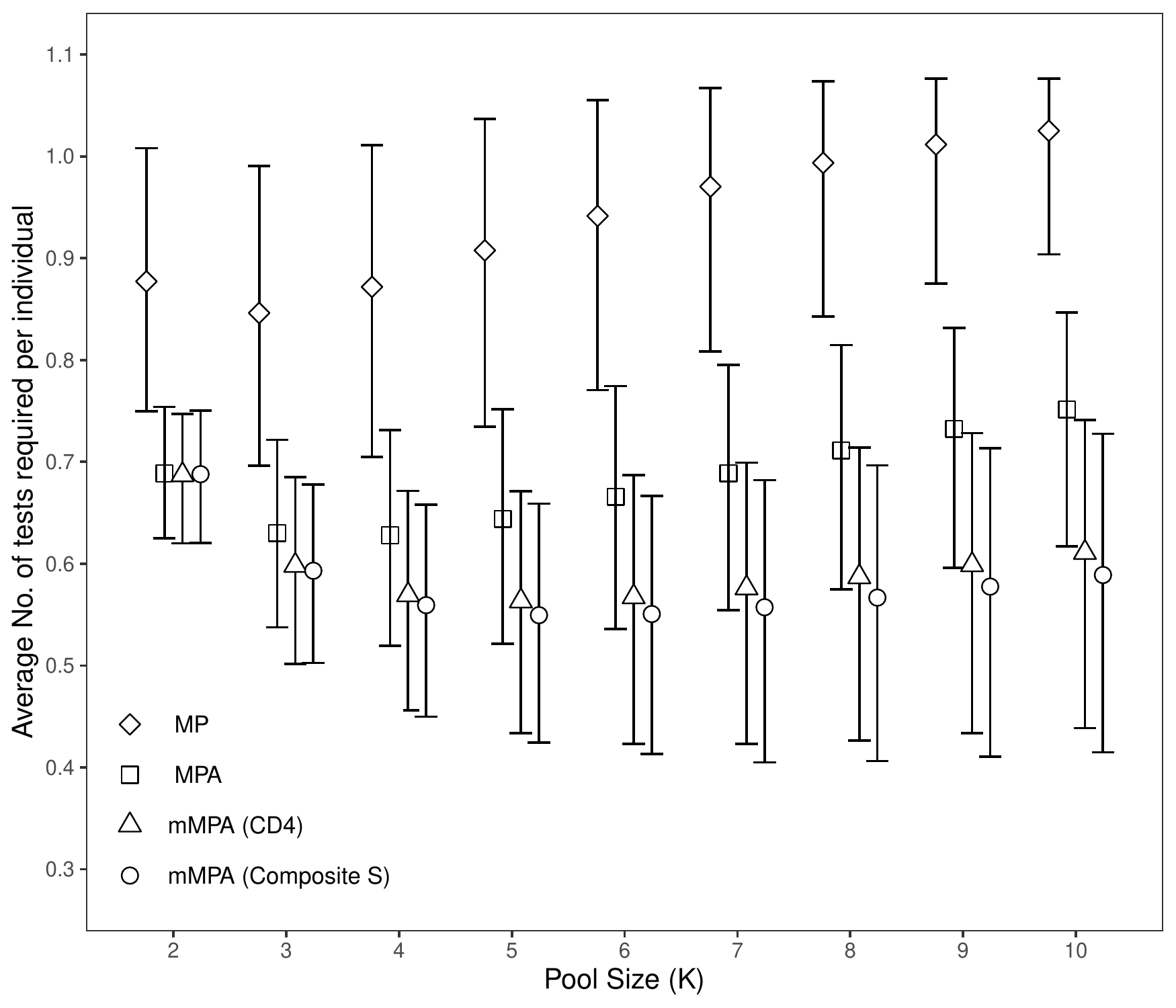}
\par\end{centering}
\caption{Estimated average number of HIV VL tests needed per individual for MP, MPA, mMPA (using CD4 count as risk score) and mMPA (using a composite  risk score) for pool sizes $K=2, \dots, 10$. The bars show  95\% confidence intervals obtained using the bootstrap method with 1,000 resamples. \label{fig:Estimated-phi}}

\end{figure}

\begin{figure}
\begin{centering}
\includegraphics[width=5.5in]{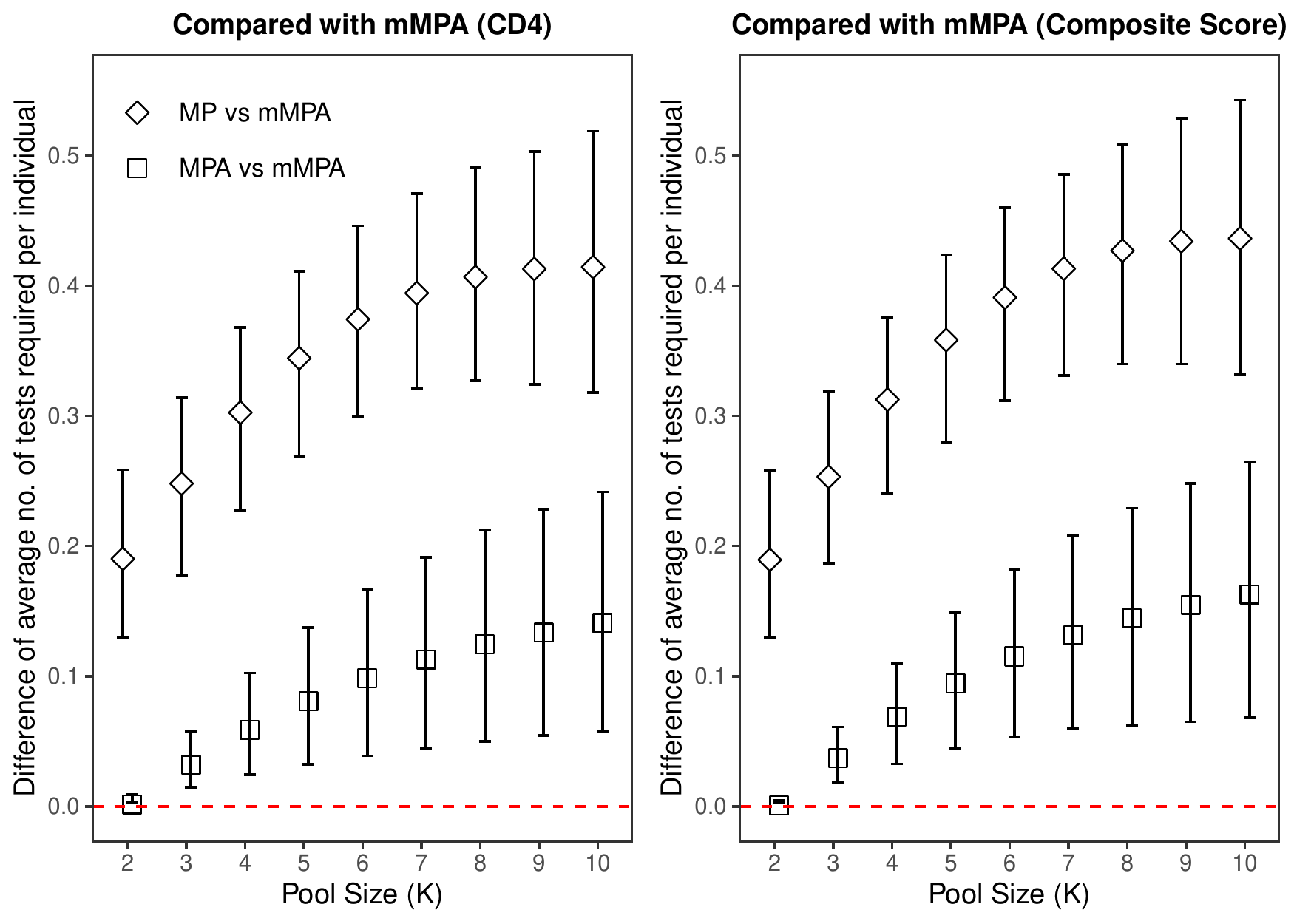}
\par\end{centering}
\caption{ Comparison of testing efficiencies between MP and mMPA (``$\Diamond$'') and between MPA and mMPA (``$\times$''). Two risk scores were used
for mMPA: CD4 count (left panel) and a composite risk score derived
from CD4 count, CD4\%, and their 6-month changes (right panel). The
bars show 95\% confidence intervals for their differences obtained
using the bootstrap method with 1,000 resamples. \label{fig:Comparison-of-ATR}}

\end{figure}

Figure \ref{fig:Estimated-phi} shows the testing efficiencies of MP, MPA, and mMPA for pool sizes between 2 and 10. The optimal pool sizes for MP and MPA are 3 and 4, and the resulting test efficiencies are 0.846 ($0.696, 0.990$) and 0.628 ($0.519, 0.731$) assay per individual, respectively. When a large pool size is used (e.g.~$K\ge8$), MP can be less efficient and need more assays than individual testing. The testing efficiency of MPA is higher than MP but lower than mMPA. Figure \ref{fig:Comparison-of-ATR} compares MP and MPA with mMPA. The 95\% CIs of their differences are calculated using ``paired'' bootstrap samples: With each bootstrap resample, we estimate $\phi$'s for MP, MPA, and mMPA and their pairwise differences, and then calculate the bootstrapped 95\% CIs based on the pairwise differences. As shown in Figure \ref{fig:Comparison-of-ATR}, the testing efficiency of mMPA is significantly higher than those of MPA and MP even though in Figure \ref{fig:Estimated-phi} their corresponding 95\% CIs overlap.  

\section{Discussion \label{sec:Conclusion}}

Pooled testing methods with quantitative assays (MP, MPA, and mMPA) can improve testing efficiency by reducing the cost in testing assays and are valuable particularly when disease/failure prevalence is low. Among all these pooling strategies, MP is the most straightforward to implement. MPA uses a deconvolution algorithm to accelerate the process of ascertaining individual failure statuses in a positive pool. mMPA improves upon MPA and uses information from other available markers to rank-order individual samples contributing to the pool, resulting in faster and more efficient deconvolution and fewer tests than MP and MPA. 

These pooling strategies can be applied in a similar fashion as pooled testing with qualitative assays, but have different statistical properties. In this paper, we formulate the pooling procedures of MP, MPA, and mMPA. We derive and compare their testing efficiencies in terms of the average number of assays required per individual, and establish the superiority of mMPA to MPA and MP. We investigate the impact of different choices of risk scores on mMPA and derive the testing efficiency bound of the methods. We also investigate the diagnostic accuracy of the three methods compared to individual testing. 

Our results on MPA and mMPA are derived based on a convolution of quantitative results. However, not all pooling with quantitative assays can be conceptualized as a convolution of individual results. For example, another commonly-used measure of virus concentration in a sample is `cycle time value' (CTV), which essentially counts the number of duplications needed for the virus can be detected by machine. Although CTV can be transformed back to virus concentration, the deconvolution algorithms used by MPA and mMPA cannot be directly applied to CTV. 
%Moreover, our results do not directly apply to quantitative assays that yield negative values. 

Implementation of mMPA relies on a chosen risk score to determine an optimal order of individual testing. For a better testing efficiency and diagnostic accuracy, `strong' risk scores are generally preferred. We proposed to use Spearman's correlation to evaluate the `strength' of a risk score, since the order of individual testing is based only on the rank of the risk score. However, the relationship between Spearman's correlation and testing efficiency is not described analytically, and can depend on factors such as  failure prevalence and test result distribution. For this reason, the testing efficiency and accuracy of mMPA on one population may not be directly generalizable to another population, even though the same risk score is used. In practice, given a previously developed risk score from a similar setting, a small validation sample of new/pilot data to pool may still be necessary to assess how strong the risk score is for implementation in the current population. 

To assist clinical researchers to assess the impact of risk score and other factors (failure prevalence, pool size, etc) for implementing mMPA and MPA, we developed a software package \texttt{QuantPooledTesting} using R \citep{RCoreTeam}. In Appendix \ref{subsec:Example-R-Code}, we provide a sample R code to use the \texttt{QuantPooledTesting} package. 

Our study demonstrates that pooled testing with quantitative assays can incorporate a sequential deconvolution algorithm and improve the testing efficiency by reducing the number of assays. Programmatic incorporation of MP, MPA, or mMPA (depending on available clinical resource and clinical context) would allow patients to get more and perhaps more frequent testing if needed. Such improvement is valuable when screening and testing is needed for a massive population and when resource and time is under constraints. For example, an important input to understanding the local dynamics of COVID-19 is population-level seroprevalence,  which is typically measured through the use of individual quantitative assays.  In settings like this, where the prevalence is low and covariates associated with the probability of having COVID antibodies are available, pooling methods such as MP, MPA and mMPA could enable local health agencies to increase the volume of testing without raising cost.

\bibliographystyle{asa}
%\bibliographystyle{vancouver}
%bibliographystyle{plainnat}
%\bibliography{pooling_main_liu}
%\bibliography{pooled_testing}
\bibliography{library}

\newpage

\appendix

\section{Appendix }
\subsection{Proof of Proposition \ref{prop:num-assay-bound}\label{sec:Proof-num-assay-bounds}}
\begin{proof}
Rewrite $Z_{j}=\|V_{j}-C\|_{0}$, where $\|\cdot\|_{0}$ denotes the $L_{0}$ norm. Using the triangle inequality,
we have
\[
    \sum_{j=1}^{K}Z_{j}=\sum_{j=1}^{K}\|V_{j}-C\|_{0}\ge\|(V_{1}-C)+\dots+(V_{K}-C)\|_{0}
    %=\|\sum_{j=1}^{K}V_{j}-KC\|_{0}
    =\|\sum_{j=1}^{K}V_{j}/K-C\|_{0},
\]
which proves the left inequality of (\ref{eq:num-assays-bounds}). Further, using the fact that $\mathbf{1}(V_{j}>C)\le\lfloor V_{j}/C\rfloor$ and the property that $\lfloor A\rfloor+\lfloor B\rfloor\le\lfloor A+B\rfloor$% where $\lfloor\cdot\rfloor$ denotes the floor operator
, we have
\[
\sum_{j=1}^{K}Z_{j}\le\sum_{j=1}^{K}\lfloor V_{j}/C\rfloor\le\lfloor\sum_{j=1}^{K}V_{j}/C\rfloor=\lfloor KV_{\mathrm{poo}l}/C\rfloor.
\]
\end{proof}

\subsection{Proof of (\ref{eq:phi.MPA})} \label{subsec:phi.mpa}

\begin{proof}
Because $T_1\ge \dots \ge T_K$, we have $\mathbf{F}_{T_1}(C) \le \dots \le \mathbf{F}_{T_K}(C) $, which implies $\sum_{i=1}^{K-1}\mathbf{F}_{T_{j}}(C) \ge (K-1) \mathbf{F}_{T_1}(C)$. 
Therefore, 
\begin{equation*}
    \phi_{\mathrm{MPA}}(K) = 1-\frac{1}{K}\sum_{j=1}^{K-1}\mathbf{F}_{T_{j}}(C)  
    \le 1-\frac{(K-1)\mathbf{F}_{T_1}(C)}{K}  
    = \frac{\mathbf{F}_{T_1}(C)}{K} + 1 - \mathbf{F}_{T_1}(C)
    \le \frac{1}{K} + \Pr(T_1>C) = \phi_\mathrm{MP}(K). 
\end{equation*}
Moreover, 
\begin{equation*}
    \phi_{\mathrm{MPA}}(K) = 1-\frac{1}{K}\sum_{j=1}^{K-1}\mathbf{F}_{T_{j}}(C) \le 1 = \phi_\mathrm{IND} \quad \quad \Longrightarrow \quad \quad  \eqref{eq:phi.MPA}. 
\end{equation*}

\end{proof}

\subsection{Derivation of (\ref{eq:expc-mmpa}) \label{subsec:Derivation-of-exp-mmpa}}

Suppose that a risk score, $S \sim  \mathbf{F}_{S}$, has a continuous support. Then, $S' = \mathbf{F}_{S}(S)$ has a uniform distribution and is equivalent to $S$ for rank ordering samples. Accordingly, the order statistic $S'_{(j)} = \mathbf{F}_{S}(S_{(j)})$ has a Beta distribution with parameters~$j$ and $(K+1-j)$. To obtain the distribution of $T_{[j]}$, we rewrite
\begin{align*}
    \Pr(T_{[j]}\le C) &= \int_{s}\Pr(T_{[j]}\le C\mid S_{(j)})\Pr(S_{(j)}=s)\mathrm{d}s \\
    &=\int_{s}\Pr(T_{[j]}\le C\mid S_{(j)}=s)\mathrm{d}\mathcal{B}_{j,(K+1-j)}\{\mathbf{F}_{S}(s)\},
\end{align*}
where $\mathcal{B}_{\alpha,\beta}(\cdot)$ denotes the Beta distribution with parameters $(\alpha,\beta)$. Further, let $A_{j}^{s}(C)=\Pr(T_{[j+1]}\le C\mid S_{(j)}=s)$. Because $S_{(j)}=s$ implies that $S_{(j')} \le s$ for all $j'>j$, we can obtain $A_{j}^{s}(C)$ sequentially by
\[
    A_{j}^{s}(C)=(A_{j+1}^{s}*\mathbf{F}_{V}^{s})(C),\quad\mathrm{for}\,j=1,\dots,K-1
\]
with $A_{K}^{s}(v)=\mathbf{F}_{V:s}(v)$. Then, we have
\begin{align}
    \Pr(T_{[j]}\le v) &=\int_{s}\Pr(T_{[j+1]}+V_{[j]}\le v \mid S_{(j)})\mathrm{d}\mathcal{B}_{j,(K+1-j)}\{\mathbf{F}_{S}(s)\} \nonumber \\
    &=\int_{s}(A_{j}^{s}*\mathbf{F}_{V|s})(v)\mathrm{d}\mathcal{B}_{j,(K+1-j)}\{\mathbf{F}_{S}(s)\}.\label{eq:appdx.PrTj}
\end{align}
Note that $T_{[j+1]}$ and $V_{[j]}$ are dependent because $V_{[j]}$ is associated with the risk score of $S_{(j)}$ and the risk scores of $V_{[j+1]}, \dots, V_{[K]}$ are less than $S_{(j)}$. However, for a given value of $S_{(j)}=s$, $T_{[j+1]}$ and $V_{[j]}$ are independent, so we can use convolution operation to calculate $\Pr(T_{[j+1]}+V_{[j]}\le v\mid S_{(j)}=s)=(A_{j}^{s}*\mathbf{F}_{V|s})(v)$.

\subsection{Proof of Theorem \ref{thm:mMPA-for-two-risk-scores}\label{sec:Proof-of-Theorem-2}}

\begin{proof}
For two risk scores $S^{(a)}$ and $S^{(b)}$, suppose that $S^{(a)}$ is a stronger risk score than $S^{(b)}$ in the sense of \eqref{def:stronger score}. Then,  
\begin{align*}
    & \quad \sum_{l=j}^{K}V_{[l]}^{(a)} \preceq\sum_{l=j}^{K}V_{[l]}^{(b)} \quad \Longrightarrow \quad \Pr(T_{[j]}^{(a)}>C) \le \Pr(T_{[j]}^{(b)}>C), \: \forall j=1,\dots, K,
\end{align*}   
which further implies 
\begin{align*}
    1 + \sum_{j=1}^{K-1}\Pr(T_{[j]}^{(a)}>C)  \le 1 + \sum_{j=1}^{K-1}\Pr(T_{[j]}^{(b)}>C) \quad 
    \Longrightarrow \quad  \eqref{thm:mMPA-for-two-risk-scores}.
\end{align*}
\end{proof}

\subsection{Proof of Corollary \ref{cor:E-M-bounds}\label{sec:Proof-of-Corollary-bound}}
\begin{proof}
The lower bound of mMPA is achieved when all failing samples are tested first after a positive pool. In this case, we need one test on the pool and $k$ additional individual tests if there are $k$ failures in the pool for $k<K$. The status of the last sample is determined by the pool result and the results of the other samples. Let $M_\mathrm{mMPA}$ denote the number of tests needed by mMPA for a pool size of $K$. 
\begin{align*}
    \mathbb{E}(M_{\mathrm{mMPA}}) & = \sum_{k=0}^K \mathbb{E}(M_{\mathrm{mMPA}} \mid \sum_{j=1}^K Z_j=k )\Pr(\sum_{j=1}^K Z_j=k)   \\
    & \ge 1+\sum_{k=0}^{K-1} k\Pr(\sum_{j=1}^K Z_j=k) + (K-1)\Pr(\sum_{j=1}^{K}Z_{j}=K)\\
    &= 1+\sum_{k=0}^{K}\{k\Pr(\sum_{j=1}^{K}Z_{j}=k)\}-\Pr(\sum_{j=1}^{K}Z_{j}=K) \\ 
    &= 1+Kp-p^{K},
\end{align*}
which proves the left inequality. The upper bound of $\mathbb{E}(M_{\mathrm{mMPA}})$ occurs when the risk score being used is the weakest score that satisfies (\ref{eq:def.risk.score}). One such case is when $S$ is independent of $V$, which implies $ (V_{1}, V_{2},\dots,V_{K}) \stackrel d = (V_{[1]}, V_{[2]},\dots,V_{[K]}) $ and $ (T_{1}, T_{2},\dots,T_{K}) \stackrel d = (T_{[1]}, T_{[2]},\dots,T_{[K]}). $
In this case, the upper bound is
\begin{align*}
    \mathbb{E}(M_{\mathrm{mMPA}}) & \le\mathbb{E}(M_{\mathrm{MPA}}) = \mathbb{E}(1+\sum_{j=1}^{K-1}\mathbf{1}\{T_{[j]}>C\})\\
     & = K - \mathbb{E}\sum_{j=1}^{K-1}\{T_{[j]} \le C\} = K - \sum_{j=1}^{K-1} \mathbf F_{T_{j}}(C), 
\end{align*}
which proves the right inequality.
\end{proof}

\subsection{Proof of of Theorem \ref{thm:The-diagnostic-accuracies} \label{appndx:diagnostic}}

\subsubsection{Sensitivity}\label{appndx:sens}

\begin{proof}
Sensitivity of individual testing is the probability of an individual testing positive given that it is true positive, 
\begin{equation*}
    \mathrm{SENS}_{\mathrm{IND}} 
    %= \Pr(V_{k}\varepsilon_{k}>C\mid V_{k}>C) 
    %= \frac{\Pr(V_{k}\varepsilon_{k}>C, V_{k}>C)}{\Pr(V_{k}>C)} 
    %= \frac{\Pr(V_{k}>C\varepsilon_{k}, V_{k}>C)}{\Pr(V_{k}>C)},
    = \Pr(\widetilde V >C \mid V>C).
\end{equation*}

Using MP, we combine the individual with others and perform a pooled testing. With loss of generality, let the individual be the $k$th sample in the pool. Suppose that there also is a measurement error in the pool test $\widetilde V_\mathrm{pool} = V_\mathrm{pool} \varepsilon_\mathrm{pool}$ with $\varepsilon_\mathrm{pool} \stackrel d= \varepsilon$. Let $\widetilde T_1 = K \widetilde V_\mathrm{pool}$ be the observed value of $T_1$. The probability of the individual testing positive is  
\begin{equation}\label{eq:sens.mp}
    \mathrm{SENS}_{\mathrm{MP}} = \Pr(\widetilde T_1>C, \widetilde V_k >C \mid V_k >C)
\end{equation}
(i.e.\ when both the pool and then the individual are positive). To compare MP and MPA, we rewrite
\begin{align*}
    \mathrm{SENS}_\mathrm{MP} &= \Pr(\widetilde V_k >C \mid V_k >C) \Pr(\widetilde T_1>C \mid \widetilde V_k >C, V_k >C)  \\
    &\le \Pr(\widetilde V_k >C \mid V_k >C) = \mathrm{SENS}_\mathrm{IND}. 
\end{align*}
Furthermore, the value of $\widetilde T_1 = \varepsilon_\mathrm{pool} \sum_{j=1}^K V_j$ does not depend on $\varepsilon_k$. So given $V_k$, $\widetilde T_1$ is independent of $\widetilde V_k$ which implies  
\begin{align}
        \mathrm{SENS}_\mathrm{MP} &= \Pr(\widetilde V_k >C \mid V_k >C) \Pr(\widetilde T_1>C \mid V_k >C) \nonumber \\
        &\ge \Pr(\widetilde V_k >C \mid V_k >C) \Pr(\widetilde V_k>C \mid V_k >C) = \mathrm{SENS}_\mathrm{IND}^2. 
\end{align}

Using MPA, we run a sequence of individual tests following an algorithm after a positive pool. The $k$th individual will be tested only when the pool and all sub-pools containing the individual are positive. Let $\pi=(\pi_1, \dots, \pi_K)$ be a permutation of $\{1, \dots, K\}$ indicating the order of testing for individuals $V_1, \dots, V_K$, i.e.\ $\pi_j=k$ indicates that $j$th tested sample is from the $k$th individual. Recall that after testing the $(j-1)$th sample, the remaining samples can be regarded as a sub-pool and its pool result is calculated as $\widetilde T_{j}=\widetilde T_1 - (\widetilde V_{\pi_1} +\dots + \widetilde V_{\pi_{j-1}})$. 
So the probability of the $k$th individual testing positive is $\sum_{j=1}^K  \Pr(\widetilde T_1>C, \dots, \widetilde T_{j}>C, \widetilde V_{\pi_j} >C \mid \pi_j=k, V_k >C) \Pr(\pi_j=k\mid V_k>C)$, which using the fact of $\Pr(\pi_j=k\mid V_k>C)=1/K$ simplifies to 
\begin{equation}\label{eq:sens.mpa}
    \mathrm{SENS}_{\mathrm{MPA}} %= \frac{1}{K} \sum_{j=1}^K  \Pr( \widetilde T_{j}>C, \widetilde V_{\pi_j} >C \mid \pi_j=k, V_k >C) 
    = \frac{1}{K} \sum_{j=1}^K  \Pr( \widetilde T_{j}>C, \widetilde V_k >C \mid \pi_j=k, V_k >C). 
\end{equation}
    
Using mMPA, we assume that for each individual $V_j$, a risk score $S_j$ is calculated. Following a positive pool, individual samples are then tested in the decreasing order of their risk scores. Let $\pi'=\{\pi_1', \dots, \pi_K'\}$ denote the resulting testing order, where $\pi_j'=k$ means that $j$th tested sample is from the $k$th individual. The values of $\pi'$s satisfy the condition $S_{\pi_1'}\ge \dots \ge S_{\pi_K'}$, i.e.\ individuals with the highest risk scores are tested first. Similar to the derivation of MPA, we can show that, 
\begin{equation} \label{eq:sens.mmpa}
    \mathrm{SENS}_{\mathrm{mMPA}} = \sum_{j=1}^K  \Pr( \widetilde T_{[j]}>C, \widetilde V_k >C \mid  \pi_j' = k , V_k >C)\Pr(\pi_j'=k \mid V_k>C),  
\end{equation}
with $\widetilde T_{[j]}=\widetilde T_1 - (\widetilde V_{\pi_1'} +\dots + \widetilde V_{\pi_{j-1}'})$. 

Because $\widetilde T_j \le \widetilde T_1$ and $\widetilde T_{[j]} \le \widetilde T_1$ for all $j=1, \dots, K$, we obtain the following order by comparing \eqref{eq:sens.mpa} and \eqref{eq:sens.mmpa} with \eqref{eq:sens.mp},
$$ \{\mathrm{SENS}_{\mathrm{MPA}}, \mathrm{SENS}_{\mathrm{mMPA}}\} \le \mathrm{SENS}_{\mathrm{MP}} \le \mathrm{SENS}_{\mathrm{IND}}. $$
The proof of relationship between $\mathrm{SENS}_{\mathrm{mMPA}}$ and $\mathrm{SENS}_{\mathrm{mMPA}}$ is less straightforward. With the Definition~\eqref{eq:def.risk.score}, we have $T_j \succeq T_{[j]}$, which implies that $\Pr( \widetilde T_{j}>C, \widetilde V_k >C \mid  V_k >C) \ge \Pr( \widetilde T_{[j]}>C, \widetilde V_k >C \mid  V_k >C)$. On the other hand, compared with $\Pr(\pi_j=k) = 1/K$ in \eqref{eq:sens.mpa}, $\Pr(\pi_j'=k \mid V_k>C)$ in \eqref{eq:sens.mmpa} is a decreasing function and places more mass on smaller values of $k$. Although we cannot theoretically prove the relationship between $\mathrm{SENS}_{\mathrm{mMPA}}$ and $\mathrm{SENS}_{\mathrm{MPA}}$, we expect that with a good risk score, mMPA can outperform MPA with the fact that at one extreme when $S \perp V$, $\mathrm{SENS}_{\mathrm{mMPA}}=\mathrm{SENS}_{\mathrm{MPA}}$, and at the other extreme when $S$ leads to a perfect rank order of $V$, $\mathrm{SENS}_{\mathrm{mMPA}}$ approaches $\mathrm{SENS}_{\mathrm{MP}}$ (when there is only one failure in a pool, $\mathrm{SENS}_{\mathrm{mMPA}}=\mathrm{SENS}_{\mathrm{MP}}$).  
\end{proof}

\subsubsection{Specificity}

\begin{proof}
The specificity of an assay is the probability of classifying an individual sample as negative given it is true negative,  
\begin{equation*}
    \mathrm{SPEC}_{\mathrm{IND}} 
    = \Pr(\widetilde V \le C \mid V\le C).
\end{equation*}

Using MP, we classify an individual (suppose that it is the $k$th sample in a pool) as negative when either the pool tests negative, or the pool tests positive but the individual sample tests negative, so
\begin{equation} \label{eq:spec.mp}
    \mathrm{SPEC}_{\mathrm{MP}} = \Pr(\widetilde T_1\le C \mid V_k \le C) + \Pr(\widetilde T_1>C, \widetilde V_k \le C \mid V_k \le C). 
\end{equation}
Because $\Pr(\widetilde T_1\le C \mid V_k \le C)  = \Pr(\widetilde T_1\le C, \widetilde V_k >C \mid V_k \le C) + \Pr(\widetilde T_1\le C, \widetilde V_k \le C \mid V_k \le C) $, we have 
\begin{align*}
    \mathrm{SPEC}_{\mathrm{MP}} = \Pr(\widetilde T_1\le C, \widetilde V_k >C \mid V_k \le C) + \Pr(\widetilde V_k \le C \mid V_k \le C) \quad \Longrightarrow \quad \mathrm{SPEC}_\mathrm{MP} \ge \mathrm{SPEC}_\mathrm{IND}. 
\end{align*}

Using MPA, we run a sequence of individual tests following a positive pool. We classify the $k$th individual as negative when (1) the pool tests negative, (2) any of the sub-pools containing the individual tests negative, or (3) all of the sub-pools containing the individual test positive but the test of the $k$th sample itself is negative. So the probability that the $k$th individual tests negatives is 
\begin{align*}
    \mathrm{SPEC}_{\mathrm{MPA}} & =  \sum_{j=1}^K \{ \Pr(\min(\widetilde T_1, \dots, \widetilde T_j) \le C \mid \pi_j=k, V_k\le C) \\
    & \quad + \Pr(\min(\widetilde T_1, \dots, \widetilde T_j) > C, \widetilde V_k \le C \mid  \pi_j=k, V_k\le C) \} \Pr(\pi_j=k \mid V_k \le C) . 
\end{align*}
Because $\widetilde T_1 \ge \widetilde T_1 \ge \dots \ge \widetilde T_j$, we have $ \min(\widetilde T_1, \dots, \widetilde T_j)  = \widetilde T_{j} $. The above equation is simplified to
%%%
\begin{align} \label{eq:spec.mpa}
    \mathrm{SPEC}_{\mathrm{MPA}} = \frac{1}{K} \sum_{j=1}^K \{ \Pr(\widetilde T_j\le C \mid \pi_j=k, V_k \le C) + \Pr(\widetilde T_j>C, \widetilde V_k \le C \mid \pi_j=k, V_k \le C) \} . 
\end{align}
Note that we can rewrite \eqref{eq:spec.mp} and \eqref{eq:spec.mpa} as 
\begin{align*}
     \mathrm{SPEC}_{\mathrm{MP}} & = 1- \Pr(\widetilde T_1>C, \widetilde V_k >C \mid V_k \le C) \\ 
     \mathrm{SPEC}_{\mathrm{MPA}} & = 1 - \frac{1}{K} \sum_{j=1}^K \Pr(\widetilde T_j>C, \widetilde V_k > C \mid \pi_j=k, V_k \le C).     
\end{align*}
It is straightforward to show that $\mathrm{SPEC}_{\mathrm{MPA}} \ge \mathrm{SPEC}_{\mathrm{MP}}$ with the fact $\widetilde T_1 \ge \widetilde T_j,\  j=1,\dots, K$ . 

Using mMPA, we test individual samples in the decreasing order of their estimated risk of failure. Using a similar approach, we can show that for mMPA, 
%%%
\begin{align} 
     \mathrm{SPEC}_{\mathrm{MPA}} = & \sum_{j=1}^K \{ \Pr(\widetilde T_{[j]}\le C \mid \pi_j'=k, V_k \le C) \nonumber \\ 
     & + \Pr(\widetilde T_{[j]}>C, \widetilde V_k \le C \mid \pi_j=k', V_k \le C) \}\Pr(\pi_j'=k \mid V_k \le C) , \label{eq:spec.mmpa}
\end{align}
and $\mathrm{SPEC}_{\mathrm{MPA}} \ge \mathrm{SPEC}_{\mathrm{MP}}$. 
\end{proof}

\subsubsection{Positive predictive value}

\begin{proof}
The positive predictive value (PPV) of an assay is the probability of an individual being true positive given that it tests positive.
\begin{equation*}
    \mathrm{PPV}_{\mathrm{IND}} 
    = \Pr(V >C \mid \widetilde V>C).
\end{equation*}

For MP, the PPV of an individual testing positive using MP is given by
\begin{equation} \label{eq:ppv.mp}
    \mathrm{PPV}_{\mathrm{MP}} = \Pr(V_k >C \mid \widetilde T_1 >C, \widetilde V_k >C),
\end{equation}
which can be expressed as 
$   \mathrm{PPV}_{\mathrm{MP}} =  \frac{\Pr(\widetilde T_1 >C \mid \widetilde V_k >C, V_{k}>C) \mathrm{PPV}_{\mathrm{IND}} } {\Pr(\widetilde T_1 >C \mid \widetilde V_{k}>C)} $. 
%% the following statement is correct but not needed. 
%Because $\widetilde T_1$ does not depend on $\varepsilon_k$, the numerator of the first term can be simplified as $\Pr(\widetilde T_1 >C \mid V_{k}>C)$. 
We can rewrite the denominator as 
\begin{align*}
    %&\quad \Pr(\widetilde T_1 >C \mid \widetilde V_{k}>C)  \\
    %&= 
    & \Pr(\widetilde T_1 >C \mid \widetilde V_{k}>C, V_k>C)\Pr(V_k>C \mid \widetilde V_k>C) + \Pr(\widetilde T_1 >C \mid \widetilde V_{k}>C, V_k \le C)\Pr(V_k\le C \mid \widetilde V_k>C) \\
    &\le \Pr(\widetilde T_1 >C \mid \widetilde V_{k}>C, V_k>C)\Pr(V_k>C \mid \widetilde V_k>C) +  \Pr(\widetilde T_1 >C \mid \widetilde V_{k}>C, V_k >C)\Pr(V_k\le C \mid \widetilde V_k>C)  \\
    &= \Pr(\widetilde T_1 >C \mid \widetilde V_{k}>C, V_k >C).
\end{align*}
Then, it follows immediately that $ \frac{\Pr(\widetilde T_1 >C \mid \widetilde V_k >C, V_{k}>C)} {\Pr(\widetilde T_1 >C \mid \widetilde V_{k}>C)}\ge 1$ and  
%%%%
$$\mathrm{PPV}_{\mathrm{MP}} \ge \mathrm{PPV}_{\mathrm{IND}}.$$

Using MPA, an individual tests positive only when the pool, all sub-pools before testing the individual, and the individual test of itself are all positive. So the probability of being true positive given that it tests positive is 
\begin{align} \label{eq:ppv.mpa}
    \mathrm{PPV}_{\mathrm{MPA}} %= \frac{1}{K} \sum_{j=1}^K  \Pr( \widetilde T_{j}>C, \widetilde V_{\pi_j} >C \mid \pi_j=k, V_k >C) 
   % = \frac{1}{K} \sum_{j=1}^K  \Pr( \widetilde T_{j}>C, \widetilde V_k >C \mid \pi_j=k, V_k >C).
    = \frac{\sum_{j=1}^K  \Pr( \widetilde T_{j}>C, \widetilde V_k >C, V_k >C, \pi_j=k)}{\sum_{j=1}^K  \Pr( \widetilde T_{j}>C, \widetilde V_k >C , \pi_j=k)}  %\nonumber \\
    =  \sum_{j=1}^K  \Pr( V_k >C  \mid  \widetilde T_{j}>C, \widetilde V_k >C, \pi_j=k) q_j, 
\end{align} 
where $$
    q_j %= \frac{1}{K} \sum_{j=1}^K  \Pr( \widetilde T_{j}>C, \widetilde V_{\pi_j} >C \mid \pi_j=k, V_k >C) 
    = \frac{\Pr( \widetilde T_{j}>C, V_k >C, \pi_j=k)}{ \sum_{l=1}^K  \Pr( \widetilde T_l>C, V_k >C, \pi_l=k)} 
    %= \Pr(\pi_j=k \mid \textrm{``test positive''}) 
$$
is a PMF. Note that \eqref{eq:ppv.mpa} is a weighted average of $\Pr( V_k >C  \mid  \widetilde T_{j}>C, \widetilde V_k >C, \pi_j=k)$, which we will show  as follow is greater than or equal to \eqref{eq:ppv.mp}, 
\begin{align*}
    & \quad \Pr( V_k >C  \mid  \widetilde T_1>C, \widetilde V_k >C) \\
    & = \Pr( V_k >C  \mid  \widetilde T_1>C, \widetilde T_j>C, \pi_j=k, \widetilde V_k >C) w +\Pr( V_k >C  \mid  \widetilde T_1>C, \widetilde T_j\le C, \pi_j=k, \widetilde V_k >C)(1-w) \\
    & \le \Pr( V_k >C  \mid  \widetilde T_1>C, \widetilde T_j>C, \pi_j=k, \widetilde V_k >C) w +\Pr( V_k >C  \mid  \widetilde T_1>C, \widetilde T_j>C, \pi_j=k, \widetilde V_k >C)(1-w) \\
    & = \Pr( V_k >C  \mid  \widetilde T_1>C, \widetilde T_j>C, \pi_j=k, \widetilde V_k >C), 
\end{align*}
where $w=\Pr(\widetilde T_j>C, \pi_j=k \mid \widetilde T_1>C, \widetilde V_k >C)$. With the above result, we obtain that $$ \mathrm{PPV}_{\mathrm{MPA}} \ge \mathrm{PPV}_{\mathrm{MP}}.  $$

Similarly for mMPA, we have 
\begin{equation} \label{eq:ppv.mmpa}
 \mathrm{PPV}_{\mathrm{mMPA}} =  \sum_{j=1}^K  \Pr( V_k >C  \mid  \widetilde T_{[j]}>C, \widetilde V_k >C, \pi_j'=k) q_j', 
\end{equation}
where 
$$q_j' = 
    \frac{\Pr( \widetilde T_{[j]}>C, V_k >C, \pi_j'=k)}{ \sum_{l=1}^K  \Pr( \widetilde T_{[l]}>C, V_k >C, \pi_l'=k)} 
    %= \Pr(\pi_j'=k \mid \textrm{``test positive''})
    . 
$$
Using the same strategy as above, we show that 
$$ \mathrm{PPV}_{\mathrm{mMPA}} \ge \mathrm{PPV}_{\mathrm{MP}}. $$
\end{proof}

\subsubsection{Negative predictive value } \label{sec:diagnosis.NPV}

\begin{proof}
The negative predictive value (NPV) of an assay is the probability of an individual being true negative given that it tests negative.
\begin{equation} \label{eq:npv.ind}
    \mathrm{NPV}_{\mathrm{IND}} 
    = \Pr(V \le C \mid \widetilde V\le C).
\end{equation}

For MP, the NPV of an individual testing negative is 
\begin{align}
    \mathrm{NPV}_{\mathrm{MP}} &=\frac{\Pr(\widetilde T_1 \le C, V_{k}\le C)+\Pr(\widetilde T_1 >C,\widetilde V_k\le C, V_{k}\le C) }{\Pr(\widetilde T_1 \le C )+\Pr(\widetilde T_1 >C, \widetilde V_k \le C)}.     \label{eq:npv.mp} 
\end{align}
To compare $\mathrm{NPV}_{\mathrm{IND}}$ and $\mathrm{NPV}_{\mathrm{MP}}$, we rewrite \eqref{eq:npv.mp} as 
\begin{align*}
    \mathrm{NPV}_{\mathrm{MP}} &=\frac{\Pr(\widetilde T_1 \le C, \widetilde V_k>C, V_{k}\le C)+\Pr(\widetilde T_1 \le C, \widetilde V_k\le C, V_{k}\le C) + \Pr(\widetilde T_1 >C,\widetilde V_k\le C, V_{k}\le C) }{\Pr(\widetilde T_1 \le C, \widetilde V_k>C ) + \Pr(\widetilde T_1 \le C, \widetilde V_k\le C) +\Pr(\widetilde T_1 >C, \widetilde V_k \le C)} \\ 
    %& =\frac{\Pr(\widetilde T_1 \le C, \widetilde V_k>C, V_{k}\le C)+\Pr(\widetilde V_k \le C, V_{k}\le C) }{\Pr(\widetilde T_1 \le C, \widetilde V_k>C ) + \Pr( \widetilde V_k \le C)}= \frac{A_1+A_2}{B_1+B_2}, 
    & = \frac{A_1+A_2+A_3}{B_1+B_2+B_3}, 
\end{align*}
where $A_1$, $A_2$, and $A_3$ denote the corresponding terms in the numerator, and $B_1$, $B_2$, and $B_3$ the corresponding terms in the denominator. Note that $A_1/B_1 = Pr(V_k\le C \mid \widetilde T_1\le C, \widetilde V_k>C)$, $A_2/B_2 = Pr(V_k\le C \mid \widetilde T_1\le C, \widetilde V_k\le C)$, and $A_3/B_3 = Pr(V_k\le C \mid \widetilde T_1>C, \widetilde V_k\le C)$. It is easy to verify that $A_1/B_1 \le A_2/B_2$. 
We further assume $A_1/B_1 \le A_3/B_3$ or more specifically, 
\begin{equation} \label{eq:NPR.assumption}
    Pr(V_k\le C \mid \widetilde T_1\le C, \widetilde V_k>C) \le Pr(V_k\le C \mid \widetilde T_1>C, \widetilde V_k\le C),
\end{equation}
which says that regardless of the pool result, if we conduct a test on an individual, the individual test result contains more information about the true status of the individual. Then we have 
\begin{align*}
    \frac{A_1}{B_1} \le \frac{A_3}{B_3} \le \frac{A_2}{B_2}   
    ~ \Longrightarrow ~ \frac{A_1}{B_2} \le \frac{A_2+A_3}{B_2+B_3} 
    ~ \Longrightarrow ~ \frac{A_1+A_2+A_3}{B_1+B_2+B_3} \le \frac{A_2+A_3}{B_2+B_3} 
    ~ \Longrightarrow ~ \mathrm{NPV}_{\mathrm{MP}} \le \mathrm{NPV}_{\mathrm{IND}}.
\end{align*}

For MPA, the NPV of an individual testing negative is given by 
\begin{equation} \label{eq:npv.mpa}
    \mathrm{NPV}_{\mathrm{MPA}} = \frac{\sum_{j=1}^K \{ \Pr(\widetilde T_j \le C, V_{k}\le C \mid \pi_j = k)+\Pr(\widetilde T_j >C, \widetilde V_k\le C, V_{k}\le C \mid \pi_j = k) \}}
    {\sum_{j=1}^K \{\Pr(\widetilde T_j \le C \mid \pi_j = k)+\Pr(\widetilde T_j >C, \widetilde V_k \le C \mid \pi_j = k) \}}. 
\end{equation}
To compare MPA and MP, we write  
\begin{align*} 
    & \Pr(\widetilde T_j \le C, V_{k}\le C \mid \pi_j = k) = \Pr(\widetilde T_j \le C, \widetilde V_k>C,  V_{k}\le C \mid \pi_j = k) + \Pr(\widetilde T_j \le C, \widetilde V_k\le C, V_{k}\le C \mid \pi_j = k)\\
    & \hspace{1in} =
    A_1 + \Pr(\widetilde T_1 >C, \widetilde T_j \le C, \widetilde V_k>C,  V_{k}\le C \mid \pi_j = k) + \Pr(\widetilde T_j \le C, \widetilde V_k\le C, V_{k}\le C \mid \pi_j = k)\\
    &\Pr(\widetilde T_j \le C \mid \pi_j = k) = \Pr(\widetilde T_j \le C, \widetilde V_{k}>C \mid \pi_j = k)  + \Pr(\widetilde T_j \le C, \widetilde V_{k}\le C \mid \pi_j = k) \\
    &\hspace{1in}= B_1 +\Pr(\widetilde T_1 >C, \widetilde T_j \le C, \widetilde V_k>C \mid \pi_j = k) + \Pr(\widetilde T_j \le C, \widetilde V_{k}\le C \mid \pi_j = k).   
\end{align*}
With some algebra, we can simplify \eqref{eq:npv.mpa} to 
\begin{align*}
    &\mathrm{NPV}_{\mathrm{MPA}} = \frac{\frac{1}{K}\sum_{j=2}^K  \Pr(\widetilde T_1>C, \widetilde T_j \le C, \widetilde V_k>C,  V_{k}\le C \mid \pi_j = k)+A_1+A_2+A_3} {\frac{1}{K}\sum_{j=2}^K \Pr(\widetilde T_1>C, \widetilde T_j  \le C, \widetilde V_k>C \mid \pi_j = k) +B_1+B_2+B_3}.
\end{align*}
Because $ \Pr( V_{k}\le C \mid \widetilde T_1>C, \widetilde T_j \le C, \widetilde V_k>C,  \pi_j = k) \le \Pr( V_{k}\le C \mid \widetilde T_1 \le C, \widetilde V_k>C) = \frac{A_1}{B_1}$, for $j=2, \dots, K$, we have 
\begin{align*}
     & \frac{ \Pr(\widetilde T_1>C, \widetilde T_j \le C, \widetilde V_k>C,  V_{k}\le C \mid \pi_j = k)} { \Pr(\widetilde T_1>C, \widetilde T_j  \le C, \widetilde V_k>C \mid \pi_j = k) }\le \frac{A_1}{B_1} \le \frac{A_3}{B_3} \le \frac{A_2}{B_2} \\
     \Longrightarrow \quad & \frac{\frac{1}{K}\sum_{j=2}^K  \Pr(\widetilde T_1>C, \widetilde T_j \le C, \widetilde V_k>C,  V_{k}\le C \mid \pi_j = k)} {\frac{1}{K}\sum_{j=2}^K \Pr(\widetilde T_1>C, \widetilde T_j  \le C, \widetilde V_k>C \mid \pi_j = k) }\le \frac{A_1+A_2+A_3 }{B_1+B_2+B_3 } \\
     \Longrightarrow \quad & \frac{\frac{1}{K}\sum_{j=2}^K  \Pr(\widetilde T_1>C, \widetilde T_j \le C, \widetilde V_k>C,  V_{k}\le C \mid \pi_j = k)+ A_1+A_2+A_3 } {\frac{1}{K}\sum_{j=2}^K \Pr(\widetilde T_1>C, \widetilde T_j  \le C, \widetilde V_k>C \mid \pi_j = k)+ B_1+B_2+B_3 }\le \frac{A_1+A_2+A_3 }{B_1+B_2+B_3 } \\
    \Longrightarrow \quad & \mathrm{NPV}_{\mathrm{MPA}} \le \mathrm{NPV}_{\mathrm{MP}}.      
\end{align*}

For MPA, the NPV of an individual with negative result is given by 
\begin{equation*} \label{eq:npv.mmpa}
    \mathrm{NPV}_{\mathrm{mMPA}} = \frac{\sum_{j=1}^K \{ \Pr(\widetilde T_{[j]} \le C, V_{k}\le C \mid \pi_j' = k)+\Pr(\widetilde T_{[j]} > C, \widetilde V_k\le C, V_{k}\le C \mid \pi_j' = k) \}}
    {\sum_{j=1}^K \{\Pr(\widetilde T_{[j]} \le C \mid \pi_j' = k)+\Pr(\widetilde T_{[j]} >C, \widetilde V_k \le C \mid \pi_j' = k) \}}.
\end{equation*}
We can use a similar approach as above to show that 
$$ \mathrm{NPV}_{\mathrm{mMPA}} \le \mathrm{NPV}_{\mathrm{MP}}. $$
\end{proof}

%\end{document}

\subsection{Sample R Code\label{subsec:Example-R-Code}} 

\begin{verbatim}
    devtools::install_github("taotliu/QuantPooledTesting")
    library("QuantPooledTesting")
    nSamp = 100
    ## simulate test results
    vSimu = round(rexp(nSamp, rate = 0.0015))

    ## simulate scores
    wgt = 0.35
    sSimu = wgt*rank(vSimu) + (1-wgt)* rank(runif(nSamp))

    ## Estimated ATR (average number of assays required per 100 subjects) of MP, MPA and mMPA
    mp_atr(vSimu, max_K = 5)
    mpa_atr(vSimu, max_K = 55)
    mmpa_atr(vSimu, sSimu, max_K = 5)
\end{verbatim}

\end{document}